\documentclass[english,aps,prc,tightenlines,floats,floatfix]{revtex4}
\usepackage{bm}
\usepackage{graphicx}

\makeatletter
\providecommand{\tabularnewline}{\\}

\usepackage{bm}
\makeatletter
\usepackage{graphics}
\usepackage{epsf}
\usepackage{babel}

\begin{document}

\title{Electromagnetic properties of the $\Delta(1232)$ and decuplet baryons
in the self-consistent SU(3) chiral quark-soliton model}

\author{Tim Ledwig}

\email{ledwig@kph.uni-mainz.de}

\affiliation{Institut f\"ur Kernphysik, Universit\"at Mainz, D-55099 Mainz, Germany}

\author{Antonio Silva}

\email{ajsilva@fe.up.pt}

\affiliation{Centro de Fisica Computacional (CFC), Departamento de Fisica, Universidade
de Coimbra, P-3004-516 Coimbra, Portugal}

\affiliation{Faculdade de Engenharia da Universidade do Porto, P-4200-465 Porto,
Portugal}

\author{Marc Vanderhaeghen }

\email{marcvdh@kph.uni-mainz.de}

\affiliation{Institut f\"ur Kernphysik, Universit\"at Mainz, D-55099 Mainz}

\date{November 2008}

\begin{abstract}
We examine the electromagnetic properties of the $\Delta(1232)$ resonance
within the self-consistent chiral quark-soliton model. In particular
we present the $\Delta$ form factors of the vector-current $G_{E0}(Q^{2})$,
$G_{E2}(Q^{2})$ and $G_{M1}(Q^{2})$ for a momentum-transfer range
of $0\leq Q^{2}\leq1\,\mbox{GeV}^{2}$. We apply the symmetry-conserving
quantization of the soliton and take $1/N_{c}$ rotational corrections
into account. Values for the magnetic moments of all decuplet baryons
as well as for the $N-\Delta$ transition are given. Special interest
is also given to the electric quadrupole moment of the $\Delta$. 
\end{abstract}



\maketitle

\section{Introduction}

The hadron spectrum can be ordered by flavor-$SU(3)$ multiplets where
the low lying baryons are assigned to either an octet or decuplet
with spin $1/2$ and $3/2$, respectively. The main focus of this
work is the hyper-charge $+1$ state of the decuplet, the $\Delta$.
Eventhough the $\Delta$ is the first excitation of the proton and
rather isolated from other resonances, due to its short life time
many of its properties are not yet experimentally determined with
accurate precision. This is reflected in the poor experimental knowledge
of the magnetic moment of the $\Delta$ which is listed by the Particle
Data Group as $\mu_{\Delta^{++}}=3.7\sim7.5\mu_{N}$ and $\mu_{\Delta^{+}}=(2.7_{-1.3}^{+1.0}(\mbox{stat.})\pm1.5(\mbox{syst.})\pm3(\mbox{theor.}))\,\mu_{N}$,
where $\mu_{N}=e/2M_{N}$ is the nucleon magneton \cite{PDG:2006}.
The former value is extracted from the reaction $\pi^{+}p\to\pi^{+}p\gamma$,
e.g. \cite{Delta++_MDM_1,Delta++_MDM_2}, and the latter one from
the process $\gamma p\to p\pi^{0}\gamma^{\prime}$ \cite{Kotulla:Delta_mag_mom_2002}.
The study of the transition process of the nucleon to the $\Delta$
can be used to gain additional information about the $N\Delta$ system.
This process is characterized by a magnetic dipole and an electric
quadrupole transition moment which are in \cite{Tiator:NDquadrupole}
extracted as $\mu_{N\Delta}=3.46\pm0.03\,\mu_{N}$ and $Q_{N\Delta}=-(0.0846\pm0.0033)\, e\mbox{fm}^{2}$,
respectively. Appart from the $\Delta$, experimental data on electromagnetic
properties of decuplet baryons only exist for the magnetic moment
of the $\Omega^{-}$ baryon $\mu_{\Omega^{-}}=(-2.02\pm0.05)\,\mu_{N}$
\cite{PDG:2006}.\\
 On the theoretical side, the $\Delta$ was investigated within many
different frameworks. In the case of $SU(6)$ symmetry the $\Delta$
magnetic moment is predicted to be $\mu_{\Delta}=Q_{\Delta}\mu_{p}$,
with $Q_{\Delta}$ being the charge of the $\Delta$ and $\mu_{p}$
the magnetic moment of the proton, which yields a value of $\mu_{\Delta^{++}}=5.58\,\mu_{N}$
\cite{SU(6)_decuplet_mag_Mom}. Other approaches include quark models
\cite{Berger:Delta_mag_Mom_CQM,Buchmann:EM_properties_D,Hong:10Mag_Mom_NRQM,Linde:Delta_Mag_Mom_CQM,Schlumpf:10Mag_Mom_QM,Ramalho:Delta_swave,Ramalho:ND_dwave},
large $N_{c}$ and soliton models \cite{Lebed:LargeNC_delta_MagMom,Kim:Delta_skyrme,Hashimoto:soltion_delta_mag_mom},
lattice QCD calculations \cite{Lee:Lattice_mag_mom_delta,Leinweber:Lattice_decuplet,Colet:lattice_Delta,Aubin:FiniteVstudy},
QCD sum rules and chiral perturbation theory \cite{Lee:QCD_sumRU_10_mag_mom,Aliev:QCD_sumRU_10mag_mom,Arndt:CHPT_10mag_mom,Banerjee:CHPT_10mag_mom,Butler:CHPT_10mag_mom,Hacker:CHPT_10magmom}.
Very recently lattice QCD calculations of electromagnetic form factors
of the $\Delta$ up to a momentum-transfer of $Q^{2}\leq2.5\,\mbox{GeV}^{2}$
were presented in \cite{Lattice_Delta_2008}. In addition, large
$N_{c}$ relations which connect the magnetic moments of the octet
and the electric quadrupole moments of the $N\Delta$ transition to
the moments of the $\Delta$ are found in \cite{Large_NC_quadrupole,LargeNc_Mag_Mom,Pascalutsa:LargeNC_delta}.\\
 In the present work we investigate the electromagnetic form factors
of the $\Delta^{+}(1232)$ in the framework of the self-consistent
chiral quark-soliton model ($\chi$QSM) assuming iso-spin symmetry.
In particular we calculate the charge ($G_{E0}$), electric quadrupole
($G_{E2}$) and magnetic dipole ($G_{M1}$) form factors of the $\Delta^{+}$
up to a momentum-transfer of $0\leq Q^{2}\leq1\,\mbox{GeV}^{2}$.
We also present values for the magnetic moments of all decuplet baryons
as well as for the $N-\Delta$ transition. In the $\chi$QSM baryons
are seen as certain $SU(3)$ rotations of a classical soliton, having
therefore the same origin. The quantization of these rotations allows
only $SU(3)$ multiplets with zero triality, hence the octet and decuplet
appear naturally. Because of this, the $\chi$QSM is able to describe
various observables of various baryons within the same set of parameters.
These parameters are fixed by reproducing mesonic experimental data,
letting the constituent quark mass to be the only free paramter in
the baryon sector. Since we can not take an exact form of the momentum-dependent
constituent quark mass we use the value of $M=420\,\textrm{MeV}$
which is known to reproduce very well the experimental data \cite{Christov:eleff,Christov:1995vm,Kim:eleff,Silva:AxialFF,Silva:ParityViol}.
The regularization behavior of the momentum-dependence is mimicked
by the proper-time regularization. The cut-off parameter and the averaged
current quark mass are then fixed for a given $M$ to the pion decay
constant $f_{\pi}$ and $m_{\pi}$, respectively. The model parameters
used in the present work are the same as in previous works \cite{AxialTheta,CQSM:ADEFF,CQSM:SHD,CQSM:Theta_VFF,Silva:AxialFF,Silva:G0A4Happex,Silva:ND,Silva:ParityViol,Silva:StrangeFF},
no additional readjusting for different observables were done. Given
that, the $\chi$QSM, with model-parameters fixed in the meson-sector
and natural inclusion of octet and decuplet baryons, provides a unique
framework with predictive power.\\
 In the past the $\chi$QSM was applied successfully to the octet
baryon (axial) vector form factors \cite{Silva:StrangeFF,Silva:ParityViol,Silva:G0A4Happex,Silva:AxialFF,Kim:eleff,Christov:eleff,CQSM:SHD},
parton- and antiparton-distributions \cite{Dressler:2001,GoekePPSU:2001,OssmannPSUG:2005,SchweitzerUPWPG:2001,Wakamatsu:2000,Wakamatsu:2005,WakamatsuN:2006,Wakamatsuref:2003}.
Furthermore, the $\chi$QSM was also applied to observables of the
anti-decuplet pentaquarks \cite{CQSM:Theta_VFF,CQSM:ADEFF,MagMom_model_indep,AxialTheta,Diakonov:1997mm,IMF_diakonov,IMF_imp_width,modelindep_theta_width}.
The vector current of decuplet baryons at $Q^{2}=0$ were investigated
in various versions of the $\chi$QSM in the past: in the self-consistent
$\chi$QSM \cite{DecupletKim,Wakamatsu:Decuplet}, in the $\chi$QSM
version formulated in the infinite momentum frame \cite{Lorce:IMF_vector}
and in the so-called \emph{model independent} $\chi$QSM version \cite{MagMom_model_indep}.
Both self-consistent $\chi$QSM calculations in the literature, which
presented the decuplet magnetic moments, were prior to the symmetry-conserving
quantization of the $\chi$QSM \cite{Praszalowicz:1998jm} which
is explicitly applied in this work and ensures the realization of
the Gell-Mann-Nishijima relation in the model.\\
 The outline of this work is as follows. In the section II we give
the general, model-independent expressions for the observables in
question. The given formulae at the end of this section are suitable
for calculation in the $\chi$QSM. Section III then describes how
these expressions are treated in the model. Final results for the
self-consistent $\chi$QSM are given in section IV. We summarize the
work in section V and give more detailed expressions in the appendix.

\section{General Formalism}

Our aim is to investigate the $\Delta(1232)$ electromagnetic form
factors and compare them to nucleon electromagnetic form factors and
the $N-\Delta$ magnetic transition moment in the self-consistent
$SU(3)$ $\chi$QSM. For that, we will summarize in this section the
relevant model-independent definitions of these quantities. The form
factors are defined through the baryon matrix-element of the vector-current
where the virtual photon couples to the $NN$, $N\Delta$ and $\Delta\Delta$
systems.

\subsection{The $\gamma^{*}NN$ Vertex}

The baryon matrix element of the vector-current, $V^{\mu\chi}(0)=\overline{\Psi}(0)\gamma^{\mu}\Psi(0)$
, between nucleon states is parametrized by two form factors $F_{1}(Q^{2})$
and $F_{2}(Q^{2})$\begin{equation}
\langle N(p^{\prime},s^{\prime})|V^{\mu}(0)|N(p,s)\rangle=\overline{u}(p,s)\Big[F_{1}(Q^{2})\gamma^{\mu}+iF_{2}(Q^{2})\frac{\sigma^{\mu\beta}q_{\beta}}{2M_{N}}\Big]u(p,s)\,\,\,,\end{equation}
 with $q=p^{\prime}-p$, $Q^{2}=-q^{2}$, $u(p,s)$ as the nucleon-spinor
of mass $M_{N}$ and third-spin component $s$. In the Breit-frame
the Sachs form factors are defined as \begin{eqnarray}
G_{E}(Q^{2})=F_{1}(Q^{2})-\frac{Q^{2}}{4M_{N}^{2}}F_{2}(Q^{2}) & \,\,\,\,\,\,\,\,\,\,\,\,\,\,\,\,\,\,\,\,\,\,\,\,\,\,\,\,\, & G_{M}(Q^{2})=F_{1}(Q^{2})+F_{2}(Q^{2})\,\,\,,\label{GE and GM normal}\end{eqnarray}
 which are projected out by the operations

\begin{eqnarray}
G_{E}(Q^{2}) & = & \int\frac{d\Omega_{q}}{4\pi}\langle N(p^{\prime},\frac{1}{2})|V^{0}(0)|N(p,\frac{1}{2})\rangle\,\,\,,\label{GE}\\
G_{M}(Q^{2}) & = & 3M_{N}\int\frac{d\Omega_{q}}{4\pi}\frac{q^{i}\epsilon^{ik3}}{i\mid\vec{q}^{\,2}\mid^{2}}\langle N(p^{\prime},\frac{1}{2})|V^{k}(0)|N(p,\frac{1}{2})\rangle\,\,\,,\label{GM}\end{eqnarray}
 where we have in the Breit-frame $Q^{2}=\vec{q}^{\,2}$. The right-hand
side of these equations can be evaluated in the $\chi$QSM.

\subsection{The $\gamma^{*}N\Delta$ Vertex}

We take the rest-frame of the final $\Delta$ with momentum $p^{\prime}=(M_{\Delta},0)$
and mass $M_{\Delta}$. The incoming nucleon has the momentum $p=(E_{N},-\vec{q})$
and energy $E_{N}$. For the $\gamma^{*}N\Delta$-Vertex we use the
decomposition of \cite{ND_vectorTransition,EM_D(1232)_excitation}.
The baryon-matrix element is written by using the Rarita-Schwinger
spinors $u^{\alpha}(p,s)$ for the $\Delta$ as \begin{eqnarray}
\langle\Delta(p^{\prime},\frac{1}{2})|V_{\mu}(0)|N(p,\frac{1}{2})\rangle & = & i\sqrt{\frac{2}{3}}\,\overline{u}^{\beta}(p^{\prime},\frac{1}{2})\,\,\Gamma_{\beta\mu}\,\, u(p,\frac{1}{2})\\
\Gamma_{\beta\mu} & = & G_{M}^{N\Delta}(Q^{2})\mathcal{K}_{\beta\mu}^{M}+G_{E}^{N\Delta}(Q^{2})\mathcal{K}_{\beta\mu}^{E}+G_{C}^{N\Delta}(Q^{2})\mathcal{K}_{\beta\mu}^{C}\,\,\,,\end{eqnarray}
 with the magnetic dipole ($G_{M}^{N\Delta}$), electric quadrupole
($G_{E}^{N\Delta}$) and Coulomb quadrupole $(G_{C}^{N\Delta})$ form
factors. The corresponding structures are \begin{eqnarray}
\mathcal{K}_{\beta\mu}^{M} & = & \frac{-3(M_{\Delta}+M_{N})}{[(M_{\Delta}+M_{N})^{2}+Q^{2}]2M_{N}}\epsilon_{\beta\mu\sigma\tau}P_{\sigma}q_{\tau}\\
\mathcal{K}_{\beta\mu}^{E} & = & -\mathcal{K}_{\beta\mu}^{M}+\frac{6}{4M_{\Delta}^{2}|\vec{q}|^{2}}\epsilon_{\beta\sigma\nu\gamma}P_{\nu}q_{\gamma}\epsilon_{\mu\sigma\alpha\delta}p_{\alpha}^{\prime}q_{\delta}i\gamma^{5}\frac{M_{\Delta}+M_{N}}{M_{N}}\\
\mathcal{K}_{\beta\mu}^{C} & = & 3\Delta^{-1}(q^{2})q_{\beta}\Big[q^{2}P_{\mu}-q\cdot Pq_{\mu}\Big]i\gamma^{5}\frac{M_{\Delta}+M_{N}}{M_{N}}\,\,\,,\end{eqnarray}
 where the momenta are defined as $P=\frac{1}{2}(p^{\prime}+p)$,
$q=p^{\prime}-p$ and $\Delta^{-1}(q^{2})=4M_{\Delta}^{2}|\vec{q}|^{2}$.
We are interested in the magnetic transition moment of the $N\to\Delta$
process. We will use again the projector $3\int\frac{d\Omega_{q}}{4\pi}\frac{q^{i}\epsilon^{ik3}}{i\mid\vec{q}\mid^{2}}$
for which the term $\mathcal{K}_{\beta k}^{C}$ vanishes. \\
 Applying the projector on the baryon matrix-element leads to\begin{eqnarray}
 &  & 3\int\frac{d\Omega}{4\pi}\frac{q^{i}\epsilon^{ik3}}{i|\vec{q}^{2}|}\langle\Delta(p^{\prime},\frac{1}{2})|V_{k}(0)|N(p,\frac{1}{2})\rangle\nonumber \\
 & = & \sqrt{\frac{E_{N}+M_{N}}{2M_{N}}}2(M_{\Delta}+M_{N})\Big[\frac{M_{\Delta}}{M_{N}}\frac{G_{M}^{N\Delta}(Q^{2})-G_{E}^{N\Delta}(Q^{2})}{[(M_{\Delta}+M_{N})^{2}+Q^{2}]}+\frac{1}{2M_{N}}\frac{G_{E}^{N\Delta}(Q^{2})}{E_{n}+M_{N}}\Big]\,\,\,.\end{eqnarray}
 The electromagnetic $N\to\Delta$ transition is dominated by the
form factor $G_{M}^{N\Delta}$ (exp. $G_{E}^{N\Delta}/G_{M}^{N\Delta}=(-2.5\pm0.5)\%$
\cite{PDG:2006}. Neglecting the $G_{E}^{N\Delta}(Q^{2})$ contribution
and taking the point $Q^{2}=0$ we have

\begin{equation}
3\int\frac{d\Omega}{4\pi}\frac{q^{i}\epsilon^{ik3}}{i|\vec{q}^{2}|}\langle\Delta(p^{\prime},\frac{1}{2})|V_{k}(0)|N(p,\frac{1}{2})\rangle|_{Q^{2}=0}=\sqrt{\frac{E_{N}(0)+M_{N}}{2M_{N}}}\frac{M_{\Delta}}{M_{N}}\frac{2}{(M_{\Delta}+M_{N})}G_{M}^{N\Delta}(0)\,\,.\label{eq:NDeltaGM}\end{equation}
 The magnetic transition moment is given by \cite{EM_D(1232)_excitation}
\begin{eqnarray}
\mu_{N\Delta} & = & \sqrt{\frac{M_{\Delta}}{M_{N}}}\, G_{M}^{N\Delta}(0)\,\mu_{N}\,\,\,,\\
Q_{N\Delta} & = & -\frac{6}{M_{N}}\frac{2M_{\Delta}}{M_{\Delta}^{2}-M_{N}^{2}}\sqrt{\frac{M_{\Delta}}{M_{N}}}\, G_{E}^{N\Delta}(0)\,\,.\end{eqnarray}
Although we will denote the quadrupole moment in units of $\mbox{fm}^{2}$
in this paper, it is understood that the electric quadrupole moment
is expressed in units of $e\mbox{fm}^{2}$, with $e$ as the electric
charge.\\
These above equations can be investigated in the $\chi$QSM.\\

\subsection{The $\gamma^{*}\Delta\Delta$ Vertex}

The baryon matrix element of the vector-current, $V^{\mu}(0)=\overline{\Psi}(0)\,\gamma^{\mu}\,\Psi(0)$
, between $\Delta$-states is parametrized by four form factors

\begin{eqnarray}
\langle\Delta(p^{\prime},s^{\prime})|V^{\mu}(0)|\Delta(p,s)\rangle & = & -\overline{u}^{\alpha}(p^{\prime},s^{\prime})\Big\{\,\,\,\,\,\,\,\,\,\,\gamma^{\mu}\Big[F_{1}^{*}g_{\alpha\beta}+F_{3}^{*}\frac{q_{\alpha}q_{\beta}}{(2M_{\Delta})^{2}}\Big]\nonumber \\
 &  & +i\frac{\sigma^{\mu\nu}q_{\nu}}{2M_{\Delta}}\Big[F_{2}^{*}g_{\alpha\beta}+F_{4}^{*}\frac{q_{\alpha}q_{\beta}}{(2M_{\Delta})^{2}}\Big]\,\,\,\,\,\,\,\,\,\,\Big\}u^{\beta}(p,s)\,\,\,.\label{eq:DJD}\end{eqnarray}
 The electric charge and quadrupole form factors $G_{E0}$, $G_{E2}$
and magnetic dipole and octupole form factors $G_{M1}$,$G_{M3}$
are defined in the Breit-frame by

\begin{eqnarray}
G_{E0}(Q^{2}) & = & (1+\frac{2}{3}\tau)\Big[F_{1}^{*}-\tau F_{2}^{^{*}}\Big]-\frac{1}{3}\tau(1+\tau)\Big[F_{3}^{*}-\tau F_{4}^{*}\Big]\\
G_{E2}(Q^{2}) & = & \Big[F_{1}^{*}-\tau F_{2}^{*}\Big]-\frac{1}{2}(1+\tau)\Big[F_{3}^{*}-\tau F_{4}^{*}\Big]\\
G_{M1}(Q^{2}) & = & (1+\frac{4}{5}\tau)\Big[F_{1}^{*}+F_{2}^{*}\Big]-\frac{2}{5}\tau(1+\tau)\Big[F_{3}^{*}+F_{4}^{*}\Big]\\
G_{M3}(Q^{2}) & = & \Big[F_{1}^{*}+F_{2}^{*}\Big]-\frac{1}{2}(1+\tau)\Big[F_{3}^{*}+F_{4}^{*}\Big]\,\,\,,\end{eqnarray}
 with $\tau=\frac{Q^{2}}{4M_{\Delta}^{2}}$. We will concentrate in
this work on the form factors $G_{E0}$, $G_{E2}$ and $G_{M1}$ and
postpone the discussion on $G_{M3}$ for future work. The zeroth-component
of the matrix-element Eq.(\ref{eq:DJD}) for both $\Delta$ having
a third-spin component of $+3/2$ reads

\begin{eqnarray}
\langle\Delta(p^{\prime},\frac{3}{2})|V^{0}(0)|\Delta(p,\frac{3}{2})\rangle & = & G_{E0}(Q^{2})-\tau\frac{2}{3}\sqrt{\frac{4\pi}{5}}Y_{20}(\Omega_{q})G_{E2}(Q^{2})\,\,\,,\end{eqnarray}
 where the projections on $G_{E0}$ and $G_{E2}$ are given by\begin{eqnarray}
G_{E0}(Q^{2}) & = & \int\frac{d\Omega_{q}}{4\pi}\langle\Delta(p^{\prime},\frac{3}{2})|V^{0}(0)|\Delta(p,\frac{3}{2})\rangle\label{eq:GE0}\\
G_{E2}(Q^{2}) & = & -\int d\Omega_{q}\sqrt{\frac{5}{4\pi}}\frac{3}{2}\frac{1}{\tau}\langle\Delta(p^{\prime},\frac{3}{2})|Y_{20}^{*}(\Omega_{q})V^{0}(0)|\Delta(p,\frac{3}{2})\rangle\,\,\,.\label{eq:GE2}\end{eqnarray}
 Using the projector $3\int\frac{d\Omega}{4\pi}\frac{q^{i}\epsilon^{ik3}}{i|\vec{q}^{2}|}$
on the $\Delta$-matrix element Eq.(\ref{eq:DJD}) gives

\begin{eqnarray}
3\int\frac{d\Omega}{4\pi}\frac{q^{i}\epsilon^{ik3}}{i|\vec{q}^{2}|}\langle\Delta(p^{\prime},\frac{3}{2})|V^{k}(0)|\Delta(p,\frac{3}{2})\rangle & = & \frac{1}{M_{\Delta}}\Big[[1+\frac{4}{5}\tau][F_{1}^{*}+F_{2}^{*}]-\tau\frac{1+\tau}{2}\frac{4}{5}[F_{3}^{*}+F_{4}^{*}]\Big]\nonumber \\
 & = & \frac{1}{M_{\Delta}}G_{M1}(Q^{2})\,\,\,.\label{eq:GM1}\end{eqnarray}
 The magnetic moment of the $\Delta$ is given by \cite{EM_D(1232)_excitation}
\begin{equation}
\mu_{\Delta}=\frac{M_{N}}{M_{\Delta}}G_{M1}(0)\mu_{N}\,\,\,,\end{equation}
 and the electric quadrupole moment by \begin{equation}
Q_{\Delta}=\frac{1}{M_{\Delta}^{2}}G_{E2}(0)\,\,\,.\end{equation}
We will also denote $Q_{\Delta}$, like $Q_{N\Delta}$ in the section
before, in units of $\mbox{fm}^{2}$. The projectors which in the
nucleon case project on the electric and magnetic form factors, project
in the $\Delta$ case on the electric charge and magnetic dipole form
factors. We will investigate Eqs.(\ref{eq:GE0},\ref{eq:GE2},\ref{eq:GM1})
in the $\chi$QSM.\\

\section{\label{sec:Form-factors-in}Form factors in the Chiral Quark-Soliton
Model}

We will now briefly describe how equations like Eqs.(\ref{GE},\ref{GM},\ref{eq:NDeltaGM},\ref{eq:GE0},\ref{eq:GE2},\ref{eq:GM1})
are evaluated in the SU(3) $\chi$QSM. For details we refer to Ref.\cite{Christov:1995vm,Christov:eleff,Kim:eleff}.
The main part of the form factors come from the baryonic matrix element
\begin{equation}
\langle B^{\prime}(p^{\prime})|\mathcal{J}^{\mu\chi}(0)|B(p)\rangle=\langle B^{\prime}(p^{\prime})|\Psi^{\dagger}(0)\mathcal{O}^{\mu\chi}\Psi(0)|B(p)\rangle,\label{matelem}\end{equation}
 where the explicit form of the operator $\mathcal{J}^{\mu\chi}=\Psi^{\dagger}(0)\mathcal{O}^{\mu\chi}\Psi(0)$
($\chi$ being a flavor index) are given by the projector in question\begin{eqnarray}
\mathcal{J}^{\mu\chi} & \to & 1\,\,\,\mbox{for the rotational Hamiltonian}\\
\mathcal{J}^{\mu\chi} & \to & \int\frac{d\Omega}{4\pi}\langle B^{\prime}(p^{\prime})|\Psi^{\dagger}(0)\gamma^{0}\gamma^{0}\Psi(0)|B(p)\rangle\,\,\,\mbox{for}\,\,\, G_{E}\\
\mathcal{J}^{\mu\chi} & \to & \int d\Omega_{q}\langle B^{\prime}(p^{\prime})|\Psi^{\dagger}(0)\gamma^{0}\gamma^{0}Y_{20}^{*}(\Omega_{q})\Psi(0)|B(p)\rangle\,\,\,\mbox{for}\,\,\, G_{E2}\\
\mathcal{J}^{\mu\chi} & \to & \int\frac{d\Omega}{4\pi}\langle B^{\prime}(p^{\prime})|\Psi^{\dagger}(0)\gamma^{0}[\vec{q}\times\vec{\gamma}]_{z}\Psi(0)|B(p)\rangle\,\,\,\mbox{for}\,\,\, G_{M},\,\, G_{M}^{N\Delta},\,\, G_{M1}\,\,\,.\end{eqnarray}
 The matrix-element Eq.(\ref{matelem}) will be treated in the path-integral
formalism with the following effective partition function of the quark
and chiral fields $\Psi$ and $U(x)$, respectively: \begin{eqnarray}
\mathcal{Z}_{\mathrm{\chi QSM}} & = & \int\mathcal{D}\psi\mathcal{D}\psi^{\dagger}\mathcal{D}U\exp\left[-\int d^{4}x\Psi^{\dagger}iD(U)\Psi\right]=\int\mathcal{D}U\exp(-S_{\mathrm{eff}}[U])\,\,\,,\label{eq:part}\\
S_{\mathrm{eff}}(U) & = & -N_{c}\mathrm{Tr}\ln iD(U)\,\,\,,\label{eq:echl}\\
D(U) & = & \gamma^{4}(i\rlap{/}{\partial}-\hat{m}-MU^{\gamma_{5}})=-i\partial_{4}+h(U)-\delta m\,\,\,,\label{eq:Dirac}\\
\delta m & = & \frac{-\overline{m}+m_{s}}{3}\gamma^{4}\bm{1}+\frac{\overline{m}-m_{s}}{\sqrt{3}}\gamma^{4}\lambda^{8}=M_{1}\gamma^{4}\bm{1}+M_{8}\gamma^{4}\lambda^{8},\label{eq:deltam}\end{eqnarray}
 where the $\mathrm{Tr}$ represents the functional trace, $N_{c}$
the number of colors, $D$ the Dirac differential operator in Euclidean
space and $\hat{m}=\mathrm{diag}(\overline{m},\,\overline{m},\, m_{\mathrm{s}})=\overline{m}+\delta m$
the current quark mass matrix of the average of the up- and down-quark
mass and strange quark mass, respectively. We assume iso-spin symmetry.
The SU(3) single-quark Hamiltonian $h(U)$ is given by

\begin{eqnarray}
h(U) & = & i\gamma^{4}\gamma^{i}\partial_{i}-\gamma^{4}MU^{\gamma_{5}}-\gamma^{4}\overline{m}\,\,\,,\label{eq:diracham}\\
U^{\gamma_{5}}(x) & = & \left(\begin{array}{lr}
U_{\mathrm{SU(2)}}^{\gamma_{5}}(x) & 0\\
0 & 1\end{array}\right)\,\,\,,\label{eq:embed}\\
U_{SU(2)}^{\gamma_{5}} & = & \exp(i\gamma^{5}\tau^{i}\pi^{i}(x))=\frac{1+\gamma^{5}}{2}U_{SU(2)}+\frac{1-\gamma^{5}}{2}U_{SU(2)}^{\dagger}\,\,\,,\end{eqnarray}
 where we use Witten's embedding of the SU(2) field $U(x)_{SU(2)}=\exp(i\tau^{i}\pi^{i}(x))$
into the SU(3). The $\pi^{i}(x)$ denote the pion-fields. We use the
factor of $N_{c}$ in Eq.(\ref{eq:echl}) in the large $N_{c}$ limit
to integrate the chiral-field in Eq.(\ref{eq:part}) with the saddle-point
approximation. For that we have to find the pion field that minimizes
the action in Eq.(\ref{eq:echl}). Generally the following Ansätze
for the chiral-field $U(x)$ and the baryon state $|B\rangle$ in
Eq.(\ref{matelem}) are made:

\begin{eqnarray}
U_{\mathrm{SU2}} & = & \exp[i\gamma_{5}\hat{n}\cdot\vec{\tau}P(r)]\,\,\,\,\,\,\,\mbox{and}\,\,\,\,\,\,\,|B(p)\rangle=\lim_{x_{4}\to-\infty}\,\frac{1}{\sqrt{\mathcal{Z}}}\, e^{ip_{4}x_{4}}\,\int d^{3}\vec{x}\, e^{i\,\vec{p}\cdot\vec{x}}\, J_{B}^{\dagger}(x)\,|0\rangle\,\,\,,\label{eq:hedgehog}\\
\mbox{with} &  & J_{B}(x)=\frac{1}{N_{c}!}\,\Gamma_{B}^{b_{1}\cdots b_{N_{c}}}\,\varepsilon^{\beta_{1}\cdots\beta_{N_{c}}}\,\psi_{\beta_{1}b_{1}}(x)\cdots\psi_{\beta_{N_{c}}b_{N_{c}}}(x)\,\,\,.\end{eqnarray}
 The first equation assumes the SU(2) field $U$ to have the most
symmetric form, a hedgehog form, with the radial pion profile function
$P(r)$ while the last two take the baryon state as an Ioffe-type
current consisting of $N_{c}$ valence quarks. The matrix $\Gamma_{B}^{b_{1}...b_{N_{c}}}$
carries the hyper-charge $Y$, isospin $I,I_{3}$ and spin $J,J_{3}$
quantum numbers of the baryon and the $b_{i}$ and $\beta_{i}$ denote
the spin-flavor- and color-indices, respectively.\\
 Applying the above treatments to the baryonic matrix element Eq.(\ref{matelem})
yields: \begin{eqnarray}
\langle B_{2}(p_{2})|\mathcal{J}^{\mu\chi}(0)|B_{1}(p_{1}\rangle & = & \frac{1}{\mathcal{Z}}\lim_{T\to\infty}e^{-ip_{2}^{4}\frac{T}{2}+ip_{1}^{4}\frac{T}{2}}\int d^{3}\vec{x}^{\prime}d^{3}\vec{x}e^{i\vec{p}_{1}\cdot\vec{x}-i\vec{p}_{2}\cdot\vec{x}^{\,\prime}}\nonumber \\
 & \times & \int\mathcal{D}U\mathcal{D}\psi^{\dagger}\mathcal{D}\psi J_{B^{\prime}}\left(\frac{T}{2},\vec{x}^{\prime}\right)\mathcal{J}^{\mu\chi}(0)J_{B}^{\dagger}\left(-\frac{T}{2},\vec{x}\right)\,\,\exp{\left[-\int d^{4}x\,\psi^{\dagger}iD(U)\psi\right]}\,\,\,.\label{eq: path-integral matrix-element}\end{eqnarray}
 Finding the minimizing chiral-field configuration $U_{c}$, the soliton,
corresponds to determine its profile function $P_{c}(r)$. This is
done by setting $\mathcal{J}^{\mu\chi}(0)=1$ in Eq.(\ref{eq: path-integral matrix-element}).
For large Euclidean times, $T\to\infty$, the expression is proportional
to the nucleon correlation function from which we can obtain the $\chi$QSM
expression for the nucleon mass. Solving numerically the equation
of motion coming from $\delta S_{eff}/\delta P(r)=0$ (minimizing
the $\chi$QSM nucleon energy) in a self-consistent approach determines
the function $P_{c}(r)$.\\
 Rotations and translations of the soliton also minimize the effective
action and are written as\begin{equation}
U(\vec{x},t)=A(t)U_{c}(\vec{x}-\vec{z}(t))A^{\dagger}(t)\,\,\,,\end{equation}
 where $A(t)$ denotes a time-dependent SU(3) matrix and $\vec{z}(t)$
stands for the time-dependent translation of the center of mass of
the soliton in coordinate space. Sofar, we considered only the classical
version of the $\chi$QSM which has to be quantized. Suitable quantum
numbers are now obtained by quantizing the rotational zero-mode. A
detailed formalism can be found in Refs.\cite{Christov:1995vm,Kim:eleff}.\\
 The Dirac operator of Eq.(\ref{eq:Dirac}) written in terms of the
soliton $U_{c}$ and its zero-modes acquires the form: \begin{equation}
D(U)=T_{z(t)}A(t)\left[D(U_{c})+i\Omega(t)-\dot{T}_{z(t)}^{\dagger}T_{z(t)}-i\gamma^{4}A^{\dagger}(t)\delta mA(t)\right]T_{z(t)}^{\dagger}A^{\dagger}(t),\label{eq:D after zeromodes}\end{equation}
 where the $T_{z(t)}$ denotes the translational operator and the
$\Omega(t)$ represents the soliton angular velocity defined as \begin{equation}
\Omega=-iA^{\dagger}\dot{A}=-\frac{i}{2}\textrm{Tr}(A^{\dagger}\dot{A}\lambda^{\alpha})\lambda^{\alpha}=\frac{1}{2}\Omega_{\alpha}\lambda^{\alpha}.\end{equation}
 The standard way to proceed is to treat all three terms $\Omega(t)$,
$\dot{T}_{z(t)}^{\dagger}T_{z(t)}$ and $\delta m$ perturbatively
by assuming a slow rotating and moving soliton and by regarding $\delta m$
as a small parameter. Generally we expand Eq.(\ref{eq:D after zeromodes})
to the first order in $\Omega(t)$, $\delta m$ and to the zeroth-order
in $\dot{T}_{z(t)}^{\dagger}T_{z(t)}$.\\
 After introducing the collective baryon wave function on the level
of Eq.(\ref{eq: path-integral matrix-element}) as \begin{equation}
\psi_{(\mathcal{R}^{*};Y^{\prime}JJ_{3})}^{(\mathcal{R};YII_{3})}(A):=\lim_{T\mapsto\infty}\frac{1}{\sqrt{\mathcal{Z}}}e^{-p^{4\prime}T/2}\int d^{3}\vec{u}^{\prime}e^{i\vec{p}^{\prime}\cdot\vec{u}^{\prime}}(\Gamma_{B}^{b_{1}...b_{N_{c}}})^{*}\Pi_{l=1}^{N_{c}}\big[\varphi_{v,b_{l}}^{\dagger}(\vec{u}^{\prime})A^{\dagger}\big]\,\,\,,\end{equation}
 and expanding the occuring fermionic determinant and product of propagators
and quantizing the soliton rotation, we obtain the following collective
Hamiltonian \cite{CQSM_Quantization}: \begin{equation}
H_{coll}=H_{\mathrm{sym}}+H_{\mathrm{sb}}\,\,\,,\label{eq:Ham}\end{equation}
 where $H_{\mathrm{sym}}$ and $H_{\mathrm{sb}}$ represent the SU(3)
symmetric and symmetry-breaking parts, respectively, \begin{eqnarray}
H_{\mathrm{sym}} & = & M_{c}+\frac{1}{2I_{1}}\sum_{i=1}^{3}J_{i}J_{i}+\frac{1}{2I_{2}}\sum_{a=4}^{7}J_{a}J_{a},\\
H_{\mathrm{sb}} & = & \frac{1}{\overline{m}}M_{1}\Sigma_{SU(2)}+\alpha D_{88}^{(8)}(A)+\beta Y+\frac{\gamma}{\sqrt{3}}D_{8i}^{(8)}(A)J_{i}\,\,\,.\end{eqnarray}
 The $M_{c}$ denotes the mass of the classical soliton and $I_{i}$
and $K_{i}$ are the moments of inertia of the soliton~\cite{Christov:1995vm},
of which the corresponding expressions can be found in Ref.\cite{Blotz_mass_splittings}
explicitly. The components $J_{i}$ denote the spin generators and
$J_{a}$ correspond to the generalized $SU(3)$ spin-generators. The
$\Sigma_{\mathrm{SU(2)}}$ is the SU(2) pion-nucleon sigma term. The
$D_{88}^{(8)}(A)$ and $D_{8i}^{(8)}(A)$ stand for the SU(3) Wigner
$D$ functions in the octet representation and the $Y$ is the hypercharge
operator. The parameters $\alpha$, $\beta$, and $\gamma$ in the
symmetry-breaking Hamiltonian are \begin{equation}
\alpha=\frac{1}{\overline{m}}\frac{1}{\sqrt{3}}M_{8}\Sigma_{SU(2)}-\frac{N_{c}}{\sqrt{3}}M_{8}\frac{K_{2}}{I_{2}},\;\;\;\;\beta=M_{8}\frac{K_{2}}{I_{2}}\sqrt{3},\;\;\;\;\gamma=-2\sqrt{3}M_{8}\left(\frac{K_{1}}{I_{1}}-\frac{K_{2}}{I_{2}}\right).\end{equation}
 The collective wave-functions of the Hamiltonian in Eq.(\ref{eq:Ham})
can be found as SU(3) Wigner $D$ functions in representation $\mathcal{R}$:
\begin{equation}
\langle A|\mathcal{R},B(YII_{3},Y^{\prime}JJ_{3})\rangle=\Psi_{(\mathcal{R}^{*};Y^{\prime}JJ_{3})}^{(\mathcal{R};YII_{3})}(A)=\sqrt{\textrm{dim}(\mathcal{R})}\,(-)^{J_{3}+Y^{\prime}/2}\, D_{(Y,I,I_{3})(-Y^{\prime},J,-J_{3})}^{(\mathcal{R})*}(A).\label{eq:Wigner}\end{equation}
 The $Y'$ is related to the eighth component of the angular velocity
$\Omega$. During the quantization process $Y^{\prime}$ is constrained
to be $Y'=-N_{c}/3=-1$. In fact, this constraint allows us to have
only SU(3) representations with zero triality.\\
 The $H_{\mathrm{sb}}$ mixes the representations for the collective
baryon states and are treated by first-order perturbation by

\begin{equation}
|B_{\mathcal{R}}\rangle=|B_{\mathcal{R}}^{\mathrm{sym}}\rangle-\sum_{\mathcal{R}^{\prime}\neq\mathcal{R}}|B_{\mathcal{R}^{\prime}}\rangle\frac{\langle B_{\mathcal{R}^{\prime}}|\, H_{\textrm{sb}}\,|B_{\mathcal{R}}\rangle}{M(\mathcal{R}^{\prime})-M(\mathcal{R})}\,\,\,.\label{wfc}\end{equation}
 From this we obtain the collective wave functions for the baryon
octet and decuplet with inclusion of wave function correction proportional
to the strange quark mass as (other wave function corrections are
listed in the appendix) \begin{eqnarray}
|N_{8}\rangle & = & |8_{1/2},N\rangle+c_{10}\sqrt{5}|\overline{10}_{1/2},N\rangle+c_{27}\sqrt{6}|27_{1/2},N\rangle\,\,\,,\label{B8}\\
|\Delta_{10}\rangle & = & |10_{3/2},\Delta\rangle+a_{27}\sqrt{\frac{15}{2}}|27_{3/2},\Delta\rangle+a_{35}\frac{5}{\sqrt{14}}|35_{3/2},\Delta\rangle\,\,\,,\end{eqnarray}
 with\begin{eqnarray}
c_{\overline{10}}=-\frac{I_{2}}{15}\Big(\alpha+\frac{1}{2}\gamma\Big) & \,\,\,\,\,\,\,\,\,\,\,\,\,\,\,\,\,\, & c_{27}=-\frac{I_{2}}{25}\Big(\alpha-\frac{1}{6}\gamma\Big)\\
a_{27}=-\frac{I_{2}}{8}\Big(\alpha+\frac{5}{6}\gamma\Big) &  & a_{35}=-\frac{I_{2}}{24}\Big(\alpha-\frac{1}{2}\gamma\Big)\,\,\,.\label{mixing_coeff}\end{eqnarray}
 Turning now to the general expression Eq.(\ref{eq: path-integral matrix-element})
for a certain operator $\mathcal{J}^{\mu\chi}(0)$ which we can now
write in the form \begin{eqnarray}
\langle B^{\prime}(p^{\prime})|\psi^{\dagger}(0)\mathcal{O}^{\mu\chi}\psi(0)|B(p)\rangle & = & \int\mathcal{D}A\int d^{3}z\,\, e^{i\vec{q}\cdot\vec{z}}\Psi_{B^{\prime}}^{*}(A)\mathcal{G}^{\mu\chi}(\vec{z})\Psi_{B}(A)e^{S_{eff}}\,\,\,,\label{general model eq}\\
 & = & \int d^{3}z\,\, e^{i\vec{q}\cdot\vec{z}}\,\langle B^{\prime}|\mathcal{G}^{\mu\chi}(\vec{z})|B\rangle\,\,\,.\end{eqnarray}
 We have used again the saddle-point approximation and expanded the
Dirac operator with respect to $\Omega$ and $\delta m$ to the linear
order and $\dot{T}_{z(t)}^{\dagger}T_{z(t)}$ to the zeroth order,
everything contained in the expression $\mathcal{G}^{\mu\chi}(\vec{z})$.
The $\mathcal{D}A$ and $d^{3}z$ arise from the zero-modes due to
summing over all $U_{c}$ configurations which minimize the $\chi$QSM
action. The expression $\mathcal{G}^{\mu\chi}(\vec{z})$ contains
the specific form factor parts originating from the explicit choice
of $\mathcal{J}^{\mu\chi}(0)$. The expansion in $\Omega$ and $\delta m$
provides the following structure of the form factors in the $\chi$QSM:
\begin{eqnarray}
G_{E,M}(Q^{2}) & = & G_{E,M}^{(\Omega^{0},m_{s}^{0})}(Q^{2})+G_{E,M}^{(\Omega^{1},m_{s}^{0})}(Q^{2})+G_{E,M}^{(m_{s}^{1}),\textrm{op}}(Q^{2})+G_{E,M}^{(m_{s}^{1}),\textrm{wf}}(Q^{2})\,\,\,,\label{eq:cqsm ff}\end{eqnarray}
 where the first term corresponds to the leading order ($\Omega^{0},m_{s}^{0}$),
the second one to the first $1/N_{c}$ rotational correction ($\Omega^{1},m_{s}^{0}$),
the third to the linear $m_{s}$ corrections coming from the operator,
and the last one to the linear $m_{s}$ corrections coming from the
wave function corrections, respectively.\\
 In the $\chi$QSM Hamiltonian of Eq.(\ref{eq:diracham}) the constituent
quark mass $M$ would in general be momentum dependenet, introducing
a natural regularization-scheme for the divergent quark loops in the
model. However, the inclusion of a momentum dependent constituent
quark mass is not straight forward and in the present framework the
standard way to proceed is to take the quark mass as a free, constant
parameter and to introduce an additional regularization scheme. The
value of $M=420\,\textrm{MeV}$ is known to reproduce very well experimental
data \cite{Christov:eleff,Christov:1995vm,Kim:eleff,Silva:AxialFF,Silva:ParityViol}
together with the proper-time regularization. In the meson-sector
the cut-off parameter and the $\overline{m}$ are then fixed for a
given $M$ to the pion decay constant $f_{\pi}$ and $m_{\pi}$, respectively.
Proceeding to the baryon-sector does not include any more new parameters.
Throughout this work the strange current quark mass is fixed to $m_{\mathrm{s}}=180\textrm{MeV}$.
We want to emphasize that all these model parameters are the same
as in previous works \cite{AxialTheta,CQSM:ADEFF,CQSM:SHD,CQSM:Theta_VFF,Silva:AxialFF,Silva:G0A4Happex,Silva:ND,Silva:ParityViol,Silva:StrangeFF},
no additional readjusting for different observables were done. The
numerical results for the moments of inertia and mixing coefficients
are summarized in Tab.\ref{Numerical-values} for $M=420$ MeV. In
case of the form factors we apply the symmetry conserving quantization
as found in \cite{Praszalowicz:1998jm}. %
\begin{table}[t]

\caption{\label{Numerical-values} Moments of inertia and mixing coefficients
for $M=420\,\textrm{MeV}$.}

\begin{centering}
\begin{tabular}{ccccccccc}
\hline 
$I_{1}\,[\textrm{fm}]$  & $I_{2}\,[\textrm{fm}]$  & $K_{1}\,[\textrm{fm}]$  & $K_{2}\,[\textrm{fm}]$  & $\Sigma_{\pi N}\,[\textrm{MeV}]$  & $c_{\overline{10}}$  & $c_{27}$  & $a_{27}$  & $a_{35}$\tabularnewline
\hline 
$1.06$  & $0.48$  & $0.42$  & $0.26$  & $41$  & $0.037$  & $0.019$  & $0.074$  & $0.018$\tabularnewline
\hline
\end{tabular}
\par\end{centering}
\end{table}

\subsection{The $\gamma^{*}NN$ Vertex in the $\chi$QSM}

We now give final expressions for Eqs.(\ref{GE},\ref{GM}) evaluated
in the $\chi$QSM on the ground of Eq.(\ref{general model eq}). References
are \cite{Christov:1995vm,Christov:eleff,Kim:eleff}. The projector
contracts the Lorentz-index and an average over the momentum transfer
orientation gives raise to spherical Bessel-functions $j_{0,1}(|\vec{q}||\vec{z}|)$.
In the Breit-frame we have $Q^{2}=|\vec{q}|^{2}$. The electric and
magnetic form factors are obtained by choosing in Eq.(\ref{eq: path-integral matrix-element})
$\mathcal{J}^{\mu}(0)$ as \begin{eqnarray}
\mathcal{J}^{\mu}(0) & \stackrel{E}{\to} & \Psi^{\dagger}\,\,\gamma^{0}\gamma^{0}\,\,\Psi\,\,\,,\\
\mathcal{J}^{\mu}(0) & \stackrel{M}{\to} & \Psi^{\dagger}\,\,\gamma^{0}z^{i}\gamma^{j}\epsilon^{ij3}\,\,\Psi\,\,\,,\end{eqnarray}
 according to Eqs.(\ref{GE},\ref{GM}).\\
 The electric and magnetic form factors in the $\chi$QSM read:

\begin{eqnarray}
G_{E}(Q^{2})=\frac{1}{2}G_{E}^{\chi=3}(Q^{2})+\frac{1}{2\sqrt{3}}G_{E}^{\chi=8}(Q^{2}) & \,\,\,\,\,\,,\,\,\,\,\,\: & G_{M}(Q^{2})=\frac{1}{2}G_{M}^{\chi=3}(Q^{2})+\frac{1}{2\sqrt{3}}G_{M}^{\chi=8}(Q^{2})\label{eq:GMmodelfull}\end{eqnarray}
 with the expressions

\begin{eqnarray}
G_{E}^{\chi}(Q^{2}) & = & \int d^{3}z\, j_{0}(|\vec{q}||\vec{z}|)\,\,\int dA\langle B^{\prime}|A\rangle\mathcal{G}_{E}^{\chi}(\vec{z})\langle A|B\rangle\,\,\,,\label{eq:G_E cqsm}\\
G_{M}^{\chi}(Q^{2}) & = & M_{N}\int d^{3}z\frac{j_{1}(|\vec{q}||\vec{z}|)}{|\vec{q}|\vec{z}|}\,\,\int dA\langle B^{\prime}|A\rangle\mathcal{G}_{M}^{\chi}(\vec{z})\langle A|B\rangle\,\,\,.\label{eq:G_M cqsm}\end{eqnarray}
 The electric and magnetic densities are given by \begin{eqnarray}
\mathcal{G}_{E}^{\chi}(\vec{z}) & = & D_{\chi8}^{(8)}\sqrt{\frac{1}{3}}\mathcal{B}(\vec{z})-\frac{2}{I_{1}}D_{\chi i}^{(8)}J_{i}\mathcal{I}_{1}(\vec{z})-\frac{2}{I_{2}}D_{\chi a}^{(8)}J_{a}\mathcal{I}_{2}(\vec{z})\nonumber \\
 &  & -\frac{2}{\sqrt{3}}M_{1}D_{\chi8}^{(8)}\mathcal{C}(\vec{z})-\frac{2}{3}M_{8}D_{88}^{(8)}D_{\chi8}^{(8)}\mathcal{C}(\vec{z})\nonumber \\
 &  & +4\frac{K_{1}}{I_{1}}M_{8}D_{8i}^{(8)}D_{\chi i}^{(8)}\mathcal{I}_{1}(\vec{z})+4\frac{K_{2}}{I_{2}}M_{8}D_{8a}^{(8)}D_{\chi a}^{(8)}\mathcal{I}_{2}(\vec{z})\nonumber \\
 &  & -4M_{8}D_{8i}^{(8)}D_{\chi i}^{(8)}\mathcal{K}_{1}(\vec{z})-4M_{8}D_{\chi a}^{(8)}D_{8a}^{(8)}\mathcal{K}_{2}(\vec{z})\,\,\,,\label{eq:Ele dens}\end{eqnarray}
 and

\begin{eqnarray}
\mathcal{G}_{M}^{\chi}(\vec{z}) & = & -\sqrt{3}D_{\chi3}^{(8)}\mathcal{Q}_{0}(\vec{z})-\frac{1}{\sqrt{3}}\frac{1}{I_{1}}D_{\chi8}^{(8)}J_{3}\mathcal{X}_{1}(\vec{z})+\sqrt{3}\frac{1}{I_{1}}d_{ab3}D_{\chi b}^{(8)}J_{a}\mathcal{X}_{2}(\vec{z})+\sqrt{\frac{1}{2}}\frac{1}{I_{1}}D_{\chi3}^{(8)}\mathcal{Q}_{1}(\vec{z})\nonumber \\
 &  & +\frac{2}{\sqrt{3}}\frac{K_{1}}{I_{1}}M_{8}D_{83}^{(8)}D_{\chi8}^{(8)}\mathcal{X}_{1}(\vec{z})-2\sqrt{3}\frac{K_{2}}{I_{2}}M_{8}D_{8a}^{(8)}D_{\chi b}^{(8)}d_{ab3}\mathcal{X}_{2}(\vec{z})\nonumber \\
 &  & +2\sqrt{3}\Big[M_{1}D_{\chi3}^{(8)}+\frac{1}{\sqrt{3}}M_{8}D_{88}^{(8)}D_{\chi3}^{(8)}\Big]\mathcal{M}_{0}(\vec{z})\nonumber \\
 &  & -\frac{2}{\sqrt{3}}M_{8}D_{83}^{(8)}D_{\chi8}^{(8)}\mathcal{M}_{1}(\vec{z})+2\sqrt{3}M_{8}D_{\chi a}^{(8)}D_{8b}^{(8)}d_{ab3}\mathcal{M}_{2}(\vec{z})\,\,\,.\label{eq:Mag dens}\end{eqnarray}
 Since $M_{1}$ and $M_{8}$ are proportional to $m_{s}$ only the
first lines of the above expressions remain in case of flavor-$SU(3)$
symmetry. The expressions $\mathcal{B}(\vec{z}),...,\mathcal{M}_{2}(\vec{z})$
are given in the appendix. The Wigner D-functions depend on the rotation
$A$, e.g. $D_{\chi3}^{(\chi)}=D_{\chi3}^{(\chi)}(A)$ and expressions
like \begin{equation}
\int dA\,\langle B^{\prime}|A\rangle\, D_{\chi3}^{(8)}(A)\,\langle A|B\rangle\end{equation}
 are evaluated as described in the appendix. The value for the nucleon
mass $M_{N}$ in front of Eq.(\ref{eq:G_M cqsm}) is taken as the
value given by the classical soliton mass, i.e. by the mass of the
nucleon in the $\chi$QSM, which is by a factor of $1.36$ heavier
than the experimental mass \cite{Christov:1995vm}.

\subsection{The $\gamma^{*}N\Delta$ Vertex in the $\chi$QSM}

We now investigate Eq.(\ref{eq:NDeltaGM}) in the $\chi$QSM. In order
to evaluate the left hand side of Eq.(\ref{eq:NDeltaGM}) in the $\chi$QSM
we had to take $\lim\,\, N_{c}\to\infty$

\begin{eqnarray}
\lim_{N_{c}\to\infty}\,3\int\frac{d\Omega}{4\pi}\frac{q^{i}\epsilon^{ik3}}{i|\vec{q}^{2}|}\langle\Delta(p^{\prime},\frac{1}{2})|V_{k}(0)|N(p,\frac{1}{2})\rangle|_{Q^{2}=0} & = & \lim_{N_{c}\to\infty}\,\sqrt{\frac{E_{N}(0)+M_{N}}{2M_{N}}}2\frac{M_{\Delta}}{M_{N}}\frac{G_{M}^{N\Delta}(0)}{(M_{\Delta}+M_{N})}\,\,\,,\\
\mu_{N\Delta} & = & \lim_{N_{c}\to\infty}\,\sqrt{\frac{M_{\Delta}}{M_{N}}}\, G_{M}^{N\Delta}(0)\,\mu_{N}\,\,\,.\label{GMDN 2}\end{eqnarray}
 In the whole $\chi$QSM approach we do not take any $N_{c}^{-2}$
and also not all $N_{c}^{-1}$corrections into account, e.g. corrections
coming from the translational zero-mode in Eq.(\ref{eq:D after zeromodes})
or vabriations of the classical soliton $U_{c}$ were not considered.
According to this we could rewrite the factors of the right hand side
of Eq.(\ref{GMDN 2}) as follows:

\begin{eqnarray}
E_{N} & = & M_{N}+\frac{\vec{p}^{\,2}}{2M_{N}}+\mathcal{O}(N_{c}^{-2})\\
\sqrt{\frac{E_{N}(0)+M_{N}}{2M_{N}}} & = & 1+\mathcal{O}(N_{c}^{-2})\\
\frac{M_{\Delta}}{M_{N}} & = & \frac{M_{N}+\frac{3}{2I_{1}}}{M_{N}}=1+\frac{3}{2I_{1}M_{N}}=1+\mathcal{O}(N_{c}^{^{-2}})\\
\frac{2}{M_{\Delta}+M_{N}} & = & \frac{1}{M_{N}}\,\frac{1}{1+\frac{3}{2I_{1}M_{N}}}=\frac{1}{M_{N}}+\mathcal{O}(N_{c}^{-2})\\
\sqrt{\frac{M_{\Delta}}{M_{N}}} & = & 1+\mathcal{O}(N_{c}^{^{-2}})\,\,\,.\label{eq:largeNC1}\end{eqnarray}
 The expressions of Eq.(\ref{GMDN 2}) then reads

\begin{eqnarray}
\lim_{N_{c}\to\infty}3\int\frac{d\Omega}{4\pi}\frac{q^{i}\epsilon^{ik3}}{i|\vec{q}|^{2}}\langle\Delta(p^{\prime},\frac{1}{2})|V_{k}(0)|N(p,\frac{1}{2})\rangle|_{Q^{2}=0} & = & \frac{1}{M_{N}}G_{M}^{N\Delta}(0)\,\,\,,\\
\mu_{N\Delta} & = & G_{M}^{N\Delta}(0)\,\mu_{N}\,\,\,.\end{eqnarray}
 The $\chi$QSM expression is then given by\begin{eqnarray}
G_{M}^{N\Delta}(0) & = & \frac{1}{2}G_{M}^{N\Delta\chi=3}(0)+\frac{1}{2\sqrt{3}}G_{M}^{N\Delta\chi=8}(0)\,\,\,,\label{eq:GNDcqsm_f}\\
G_{M}^{N\Delta\chi}(0) & = & M_{N}\int d^{3}z\frac{j_{1}(|\vec{q}||\vec{z}|)}{|\vec{q}||\vec{z}|}|_{Q^{2}=0}\,\,\int dA\langle\Delta(\frac{1}{2})|A\rangle\mathcal{G}_{M}^{\chi}(\vec{z})\langle A|N(\frac{1}{2})\rangle\,\,\,,\end{eqnarray}
 where the density $\mathcal{G}_{M}^{\chi}(\vec{z})$ is the same
as in Eq.(\ref{eq:G_M cqsm}) since the projectors in Eqs.(\ref{GM},\ref{eq:NDeltaGM})
are the same. The only $1/N_{c}$ correction which is taken into account
on the level of Eq.(\ref{general model eq}) are those originating
from $\mathcal{G}(\vec{z})$ but not from the expression $e^{i\vec{q}\cdot\vec{z}}$.
This is connected to the fact that we just expand Eq.(\ref{eq:D after zeromodes})
to the zeroth-order in $\dot{T}_{z(t)}^{\dagger}T_{z(t)}$. In case
of the rest-frame of the $\Delta$ we have for $\vec{q}^{\,2}$ the
expression

\begin{eqnarray}
\vec{q}^{\,2} & = & (M_{\Delta}-E_{N})^{2}+Q^{2}=Q^{2}+\mathcal{O}(N_{c}^{-2})\\
|\vec{q}| & = & \sqrt{Q^{2}}+\mathcal{O}(N_{c}^{-2})\,\,\,.\end{eqnarray}
 This means in the present formalism the $|\vec{q}|$ entering in
Eq.(\ref{eq:GNDcqsm_f}) is actually $\sqrt{Q^{2}}$. Applying the
above large $N_{c}$ arguments means, we neglect \emph{all} $1/N_{c}$
corrections beside those coming from the rotational frequency ($\Omega$)
expansion of Eq.(\ref{eq:D after zeromodes}). After having done this,
we put $N_{c}=3$ in order to get finite numerical numbers.

\subsection{The $\gamma^{*}\Delta\Delta$ Vertex in the $\chi$QSM\label{sub:DeltaDeltaCQSM}}

For the $\Delta$ electromagnetic form factors we use again the Breit-frame
with $Q^{2}=\vec{q}^{\,2}$ and \begin{eqnarray}
G_{E0}(Q^{2}) & = & \frac{1}{2}G_{E0}^{\chi=3}(Q^{2})+\frac{1}{2\sqrt{3}}G_{E0}^{\chi=8}(Q^{2})\,\,\,,\\
G_{E2}(Q^{2}) & = & \frac{1}{2}G_{E2}^{\chi=3}(Q^{2})+\frac{1}{2\sqrt{3}}G_{E2}^{\chi=8}(Q^{2})\,\,\,,\\
G_{M1}(Q^{2}) & = & \frac{1}{2}G_{M1}^{\chi=3}(Q^{2})+\frac{1}{2\sqrt{3}}G_{M1}^{\chi=8}(Q^{2})\,\,\,.\label{eq:GM1modelfull}\end{eqnarray}
 The projector of the electric charge form factor of the $\Delta$
is the same as for the nucleon case, hence we can use Eq.(\ref{eq:G_E cqsm})
with \begin{equation}
G_{E0}^{\chi}(Q^{2})=\int d^{3}z\, j_{0}(|\vec{q}||\vec{z}|)\langle\Delta(\frac{3}{2})|\mathcal{G}_{E}^{\chi}(\vec{z})|\Delta(\frac{3}{2})\rangle\,\,\,.\label{eq:G_E0 cqsm}\end{equation}
 The $\Delta$ magnetic dipole form factor Eq.(\ref{eq:GM1}) and
magnetic moment have the pre-factors \begin{eqnarray}
\frac{1}{M_{\Delta}} & = & \frac{1}{M_{N}+\frac{3}{2I_{1}}}=\frac{1}{M_{N}}\frac{1}{1+\mathcal{O}(N_{c}^{2})}\,\,\,,\\
\frac{M_{N}}{M_{\Delta}} & = & \frac{M_{N}}{M_{N}}\frac{1}{1+\mathcal{O}(N_{c}^{2})}\,\,\,,\label{eq:largeNC2}\end{eqnarray}
 and give therefore the expressions in the $\chi$QSM

\begin{eqnarray}
G_{M1}^{\chi}(Q^{2}) & = & M_{N}\int d^{3}z\frac{j_{1}(|\vec{q}||\vec{z}|)}{|\vec{q}||\vec{z}|}\langle\Delta(\frac{3}{2})|\mathcal{G}_{M}^{\chi}(\vec{z})|\Delta(\frac{3}{2})\rangle\,\,\,,\label{eq:G_M1 cqsm}\\
\mu_{\Delta} & = & G_{M1}(0)\,\mu_{N}\,\,\,.\end{eqnarray}
 The densities $\mathcal{G}_{E}^{\chi}(\vec{z})$ and $\mathcal{G}_{M}^{\chi}(\vec{z})$
are the same as in Eqs.(\ref{eq:G_E cqsm},\ref{eq:G_M cqsm}) since
the projectors in Eqs.(\ref{GE},\ref{eq:GE0}) and Eqs.(\ref{GM},\ref{eq:GM1})
are the same, respectively.\\
 The projector on $G_{E2}$ is different. The electric quadrupole
form factor reads in terms of Eq.(\ref{general model eq}) \begin{eqnarray}
G_{E2}^{\chi}(Q^{2}) & = & -\int d\Omega_{q}\sqrt{\frac{5}{4\pi}}\frac{3}{2}\frac{1}{\tau}\int dz^{3}e^{i\vec{q}\cdot\vec{z}}\langle\Delta(\frac{3}{2})|[Y_{20}^{*}(\Omega_{q})\mathcal{G}^{0\chi}(\vec{z})]|\Delta(\frac{3}{2})\rangle\,\,\,,\end{eqnarray}
 which gives after performing the integral over $d\Omega_{q}$\begin{eqnarray}
G_{E2}^{\chi}(Q^{2}) & = & 6\sqrt{5}M_{\Delta}^{2}\int dr\,\, r^{4}\,\frac{j_{2}(k\cdot r)}{k^{2}r^{2}}\,\,\int d\Omega_{z}\langle\Delta(\frac{3}{2})|\,\,[\sqrt{4\pi}Y_{20}(\Omega_{z})\mathcal{G}^{0\chi}(\vec{z})]\,\,|\Delta(\frac{3}{2})\rangle\,\,\,,\end{eqnarray}
 with $r=|\vec{z}|$ and $k=|\vec{q}|$. The expression $[\sqrt{4\pi}Y_{20}(\Omega_{z})\mathcal{G}^{0\chi}(\vec{z})]=\mathcal{G}_{E2}^{0\chi}(\vec{z})$
shall illustrate the $\chi$QSM form factor density which we obtain
when we choose the operator $\mathcal{J}^{\mu}(0)$ in Eq.(\ref{eq: path-integral matrix-element})
as \begin{equation}
\mathcal{J}^{\mu}(0)\stackrel{E2}{\to}\Psi^{\dagger}\,\,\sqrt{4\pi}Y_{20}(\Omega_{z})\gamma^{0}\gamma^{0}\,\,\Psi\,\,\,,\end{equation}
 according to Eq.(\ref{eq:GE2}).\\
 Since $G_{E2}$ is extracted out from the zeroth-component of the
vector-current the Lorentz-structure is the same as for the form factor
$G_{E}$. Hence, we can construct the $G_{E2}$ $\chi$QSM form factor
density from the expression for $G_{E}$. For the form factor $G_{E2}$
we will not take any $m_{s}$-corrections coming from the operator
into account and start from the $SU(3)$-expression of $G_{E}$ which
reads

\begin{eqnarray}
G_{E}^{\chi}(Q^{2}) & = & \int d^{3}z\, j_{0}(|\vec{q}||\vec{z}|)\,\,\int dA\langle B^{\prime}|A\rangle\mathcal{G}_{E}^{\chi}(\vec{z})\langle A|B\rangle\,\,\,,\end{eqnarray}
 with the density \begin{eqnarray*}
\mathcal{G}_{E}^{\chi}(\vec{z}) & = & D_{\chi8}^{(8)}\sqrt{\frac{1}{3}}\mathcal{B}(\vec{z})-2\{\frac{J_{i}}{2I_{1}},D_{\chi j}^{(8)}\}\mathcal{I}_{1}^{ij}(\vec{z})-2\{\frac{J_{a}}{2I_{2}},D_{\chi a}^{(8)}\}\mathcal{I}_{2}(\vec{z})\\
\frac{1}{N_{c}}\mathcal{B}(\vec{z}) & = & \phi_{v}^{\dagger}(\vec{z})O\phi_{v}(\vec{z})-\frac{1}{2}\sum_{n}\textrm{sign}(\varepsilon_{n})\phi_{n}^{\dagger}(\vec{z})O\phi_{n}(\vec{z}),\\
\frac{1}{N_{c}}\mathcal{I}_{1}^{ij}(\vec{z}) & = & \frac{1}{2}\sum_{\varepsilon_{n}\neq\varepsilon_{v}}\frac{1}{\varepsilon_{n}-\varepsilon_{v}}\langle v|\tau^{i}|n\rangle\phi_{n}^{\dagger}(\vec{z})O\tau^{j}\phi_{v}(\vec{z})+\frac{1}{4}\sum_{n,m}\mathcal{R}_{3}(\varepsilon_{n},\varepsilon_{m})\langle n|\tau^{i}|m\rangle\phi_{m}^{\dagger}(\vec{z})O\tau^{j}\phi_{n}(\vec{z})\\
\frac{1}{N_{c}}\mathcal{I}_{2}(\vec{z}) & = & \frac{1}{4}\sum_{\varepsilon_{n^{0}}}\frac{1}{\varepsilon_{n^{0}}-\varepsilon_{v}}\langle n^{0}|v\rangle\phi_{v}^{\dagger}(\vec{z})O\phi_{n^{0}}(\vec{z})+\frac{1}{4}\sum_{n,m^{0}}\mathcal{R}_{3}(\varepsilon_{n},\varepsilon_{m^{0}})\phi_{m^{0}}^{\dagger}(\vec{z})O\phi_{n}(\vec{z})\langle n|m^{0}\rangle\,\,\,.\end{eqnarray*}
 The choice of $\mathcal{J}^{\mu}(0)$ defines the operator $O$ in
the densities $\mathcal{B},\mathcal{I}_{1},\mathcal{I}_{2}$ which
in case of the form factor $G_{E}$ is $O=\gamma^{0}\gamma^{0}=1$
and in case of $G_{E2}$ it is $O=\sqrt{4\pi}Y_{20}(\Omega_{z})$.
The density $\mathcal{B}$ originates from the zeroth-order $\Omega^{0}$
in the rotation-velocity expansion of Eq.(\ref{eq:D after zeromodes})
whereas $\mathcal{I}_{1},\mathcal{I}_{2}$ are the first rotational
$\Omega^{1}$-corrections. The $\Omega^{1}$ corrections are also
referred to as $1/N_{c}$ corrections. In case of the operator $O=\sqrt{4\pi}Y_{20}(\Omega_{z})$
the corresponding densities $\mathcal{B}(\vec{z})$ and $\mathcal{I}_{2}(\vec{z})$
are zero. \\
 The final expression in the $\chi$QSM for the form factor $G_{E2}$
reads

\begin{eqnarray}
G_{E2}^{\chi}(Q^{2}) & = & \frac{12}{I_{1}\sqrt{2}}\, M_{\Delta}^{2}\langle B^{\prime}|\Big[3D_{\chi3}^{(8)}J_{3}-D_{\chi i}^{(8)}J_{i}\Big]|B\rangle\int dr\,\, r^{4}\,\frac{j_{2}(|\vec{q}|r)}{|\vec{q}|^{2}r^{2}}\,\mathcal{I}_{1E2}(r)\,\,\,,\label{eq:GE2 CQSM Final}\end{eqnarray}
 with the density

\begin{eqnarray}
\frac{6}{N_{c}}\mathcal{I}_{1E2}(r) & = & \sum_{n\neq v}\frac{1}{\varepsilon_{n}-\varepsilon_{v}}\,(-)^{G_{m}}\,\langle A^{v},G^{v}||\tau_{1}||A^{n},G^{n}\rangle\langle A^{n},G^{n}||r\rangle\{\sqrt{4\pi}Y_{2}\otimes\tau_{1}\}_{1}\langle r||A^{v},G^{v}\rangle\\
 &  & +\frac{1}{2}\sum_{n,m}\mathcal{R}_{3}(\varepsilon_{n},\varepsilon_{m})\,(-)^{G_{n}-G_{m}}\,\langle A^{n},G^{n}||\tau_{1}||A^{m},G^{m}\rangle\langle A^{m},G^{m}||r\rangle\{\sqrt{4\pi}Y_{2}\otimes\tau_{1}\}_{1}\langle r||A^{n},G^{n}\rangle\,\,\,,\end{eqnarray}
 where the sum over the third grand-spins of the basis sates in App.\ref{sub:APP redmatelem}
are already taken. The whole $G_{E2}^{\chi}$ form factor originates
from the rotational corrections and therefore scales as $1/N_{c}$
and vanishes in the large $N_{c}$ limit.\\
 The same density also occures in the $\chi$QSM expression for the
$N-\Delta$ transition form factor ratios $R_{EM}=-G_{E}^{N\Delta}(0)/G_{M}^{N\Delta}(0)=E2/M1$
and $R_{SM}=C2/M1\sim G_{C}^{N\Delta}(0)/G_{M}^{N\Delta}(0)$ in \cite{Silva:ND}.
The final results of that $\chi$QSM $SU(3)$ analysis are $E2/M1=-1.4\%$
and $C2/M1\approx-1.8\%$ for which we can write \begin{equation}
0.78\approx\frac{E2/M1}{C2/M1}=\frac{E2}{C2}=\frac{1}{3}\frac{\int dr\,\frac{\partial}{\partial r}\Big[rj_{2}(|\vec{q}|r)\,\Big]\,\,\mathcal{I}_{1E2}(\vec{x})}{\int dr\, j_{2}(|\vec{q}|r)\,\mathcal{I}_{1E2}(\vec{x})}\,\,\,,\end{equation}
 by using the formulae presented in \cite{Silva:ND}. Inserting the
density $\mathcal{I}_{1E2}(r)$ of this work reproduces the $0.78$.
In addition we can also reproduce the values for $M^{E2}$ presented
in \cite{Watabe:ND} by using the expressions of that work with the
density $\mathcal{I}_{1E2}(r)$ of this work.

\section{Results and Discussion}

We now present and discuss the final results of this work. We have
calculated the electromagnetic form factors $G_{E0}$, $G_{E2}$ and
$G_{M1}$ of the $\Delta(1232)$ and compare them to the form factors
$G_{E}$ and $G_{M}$ of the nucleon. We also consider the magnetic
transition moment of the process $N\to\Delta$ and give numerical
values for all other decuplet magnetic moments. All results are achieved
by using the self-consistent SU(3) $\chi$QSM. In this formalism the
constituent quark mass $M$ is the only free parameter with standard
value $M=420$ MeV. Numerical parameter are fixed as described in
Sec.\ref{sec:Form-factors-in} and are exactly the same as in the
works \cite{AxialTheta,CQSM:ADEFF,CQSM:SHD,CQSM:Theta_VFF,Silva:AxialFF,Silva:G0A4Happex,Silva:ND,Silva:ParityViol,Silva:StrangeFF}.
With the numerical parameters of Tab.\ref{Numerical-values}, the
$\chi$QSM yields masses of the octet and decuplet baryons in unit
of MeV as Ref.\cite{AxialTheta}: \begin{eqnarray}
M_{N} & = & 1001(939),\,\,\,\, M_{\Lambda}=1124(1116),\,\,\,\, M_{\Sigma}=1179(1189),\,\,\,\, M_{\Xi}=1275(1318)\,\,\,,\end{eqnarray}
 \[
M_{\Delta}=1329(1232),\,\,\,\, M_{\Sigma^{*}}=1431(1385),\,\,\,\, M_{\Xi^{*}}=1533(1530),\,\,\,\, M_{\Omega}=1635(1672)\,\,\,,\]
 where the numbers in the parentheses are the experimental values
of the Particle Data Group \cite{PDG:2006}. The $\chi$QSM values
were obtained by first calculating the hyper-charge splittings with
Eq.(\ref{eq:Ham}) and afterwards starting from the experimental octet
mass center, $M_{8}=(M_{\Lambda}+M_{\Sigma})/2=1151.5$ MeV.\\
 In general for the observables investigated in this work a change
of the constituent quark mass between the values $M=(400\sim450)$
MeV affect the numerical values of the observables by $4\%$. We therefore
present only final results for $M=420$ MeV.\\
 We will first discuss the values of the form factors at the point
$Q^{2}=0$ and afterwards their $Q^{2}$ dependence up to $Q^{2}=1\,\mbox{GeV}^{2}$.\\
The magnetic moments are obtained from Eqs.(\ref{eq:GMmodelfull},\ref{eq:GNDcqsm_f},\ref{eq:GM1modelfull})
\begin{equation}
G_{M}(0)=\frac{1}{2}\Big[G_{M}^{\chi=3}(0)+\frac{1}{\sqrt{3}}G_{M}^{\chi=8}(0)\Big]\,\,\,,\end{equation}
 for which we can rewrite Eqs.(\ref{eq:Mag dens}) in the following
simple form: \begin{eqnarray}
G_{M}^{\chi}(0) & = & \int dA\,\,\langle B^{\prime}|A\rangle\Big[\hat{\mathcal{G}}_{M}^{\chi}+\hat{\mathcal{G}}_{M}^{\chi(opc)}\Big]\langle A|B\rangle\\
\hat{\mathcal{G}}_{M}^{\chi} & = & w_{1}D_{\chi3}^{(8)}+w_{2}d_{pq3}D_{\chi p}^{(8)}\hat{J_{q}}+w_{3}\frac{1}{\sqrt{3}}D_{\chi8}^{(8)}\hat{J}_{3}\\
\hat{\mathcal{G}}_{M}^{\chi(opc)} & = & w_{4}\frac{1}{\sqrt{3}}d_{pq3}D_{\chi p}^{(8)}D_{8q}^{(8)}+w_{5}(D_{\chi3}^{(8)}D_{88}^{(8)}+D_{\chi8}^{(8)}D_{83}^{(8)})+w_{6}(D_{\chi3}^{(8)}D_{88}^{(8)}-D_{\chi8}^{(8)}D_{83}^{(8)})\,\,\,.\label{GMmodel}\end{eqnarray}

\begin{table}
\begin{centering}
\begin{tabular}{cccccc}
\hline 
$w_{1}$  & $w_{2}$  & $w_{3}$  & $w_{4}$  & $w_{5}$  & $w_{6}$\tabularnewline
\hline 
$\begin{array}{c}
-12.94\,(\textrm{with}\,\mathcal{M}_{0})\\
-13.64\,(\textrm{without}\,\mathcal{M}_{0})\end{array}$  & $7.13$  & $5.16$  & $-1.31$  & $-0.78$  & $0.07$\tabularnewline
\hline
\end{tabular}
\par\end{centering}

\caption{\label{tab:Magnetic-parameter-for}Magnetic parameters for Eq.(\ref{GMmodel}).
The parameters are for a constituent quark mass of $M=420$ MeV and
a mass of $M_{N}^{\chi QSM}=939\cdot1.36$ MeV in Eq.(\ref{eq:G_M cqsm})
as described in the text. The density $\mathcal{M}_{0}$ is proportional
to $m_{s}$. }

\end{table}

All magnetic constants in this work can be reproduced (within accuracy)
by using the values of Tab.\ref{tab:Magnetic-parameter-for} and the
matrix-elements of App.\ref{sub:Matrix-Elements}. In the case of
flavor-$SU(3)$ symmetry only the paramters $w_{1}$, $w_{2}$ and
$w_{3}$ contribute whereas $w_{4}$, $w_{5}$ and $w_{6}$ are $m_{s}$
corrections coming from the operator; wave function corrections contribute
via $|B\rangle$ with the paramters $w_{1}$, $w_{2}$ and $w_{3}$.
Since the right hand-sides of Eqs.(\ref{eq:G_M cqsm},\ref{eq:GNDcqsm_f},\ref{eq:G_M1 cqsm},\ref{eq:GE2 CQSM Final})
are model-equations we also take the model-value for the nucleon mass
which is by a factor of $1.36$ larger than the experimental value,
$M_{N}^{\chi QSM}=939\cdot1.36$ MeV.\\
 As in \cite{DecupletKim} we can write the magnetic moments of the
decuplet baryons in flavor-$SU(3)$ symmetry by the simple formula\begin{equation}
\mu_{B^{10}}=-\frac{1}{12}(w_{1}-\frac{1}{2}w_{2}-\frac{1}{2}w_{3})\,\, Q_{10}\,\, J_{3}\,\,\mu_{N}\,\,\,,\end{equation}
 where $Q_{10}$ is the charge of the decuplet baryon and $J_{3}$
its third-spin component. The numerical value of this equation, given
later (Eq.(\ref{eq:decmagmomsu3})), is close to the \emph{model independent}
analysis in \cite{DecupletKim} and comparable to the one in \cite{MagMom_model_indep}.
The $\chi$QSM analysis of \cite{DecupletKim,Wakamatsu:Decuplet}
gave in flavor-$SU(3)$ a decuplet magnetic moment of $2.23\cdot Q_{10}\mu_{N}$.
Eventhough the numerical value of the present work is close, there
are differences in its determination. As explicitly mentioned in \cite{DecupletKim}
the so-called symmetry conserving quantization (SCQ) technique \cite{Praszalowicz:1998jm}
was not applied and the magnetic moment of $2.23\cdot Q_{10}\mu_{N}$
is normalized to the experimental nucleon mass in Eq.(\ref{eq:G_E cqsm}).
The SCQ has as a consequence that it decreases $\mu_{B}$, like $g_{A}^{3}$
in \cite{Silva:AxialFF} compared to \cite{Axial:Blotz} , but the
normalization to the nucleon mass as it comes out in the self-consistent
$\chi$QSM enhances $\mu_{B}$. The final numerical value for the
decuplet flavor-$SU(3)$ magnetic moment with $J_{3}=3/2$, application
of SCQ and normalization to the soliton nucleon mass, $M_{N}^{\chi QSM}=939\cdot1.36$
MeV, is \begin{equation}
\mu_{B^{10}}^{\chi QSM}=2.47\cdot Q_{10}\,\,\mu_{N}\,\,,\label{eq:decmagmomsu3}\end{equation}
 by using the values of Tab.\ref{tab:Magnetic-parameter-for}. Our
final results for the magnetic moments by including flavor-$SU(3)$
breaking effects are summarized in Tab.\ref{tab:Magnetic-Moments-of}.
The $m_{s}$ corrections of this work are more moderate compared to
the results in \cite{DecupletKim}. This is also a consequence of
the SCQ. The SCQ has a significant impact on the parameter $w_{1}$,
therefore alters the ratio of the wave function to operator corrections
in this work compared to \cite{DecupletKim}. For the wave function
corrections, numerically the factor $a_{27}$ is dominant and the
magnetic moment corrections originating from it are sensitive to $w_{1}$.
However, in general the $m_{s}$ corrections in this work are maximal
$8$\% beside the neutral baryons . The $m_{s}$ corrections in this
work have the same sign as in \cite{Wakamatsu:Decuplet} which is
not always the case by comparing with \cite{DecupletKim}.\\
\begin{table}
\begin{tabular}{c|cccccccccc}
\hline 
$\mu/\mbox{n.m.}$  & $\Delta^{++}$  & $\Delta^{+}$  & $\Delta^{0}$  & $\Delta^{-}$  & $\Sigma_{10}^{+}$  & $\Sigma_{10}^{0}$  & $\Sigma_{10}^{-}$  & $\Xi_{10}^{0}$  & $\Xi_{10}^{-}$  & $\Omega^{-}$\tabularnewline
\hline 
this work  & $4.85$  & $2.35$  & $-0.14$  & $-2.63$  & $2.47$  & $-0.02$  & $-2.52$  & $0.09$  & $-2.40$  & $-2.29$\tabularnewline
$\chi$QSM '98 \cite{DecupletKim}  & $4.73$  & $2.19$  & $-0.35$  & $-2.90$  & $2.52$  & $-0.08$  & $-2.69$  & $0.19$  & $-2.48$  & $-2.27$\tabularnewline
\hline
\end{tabular}

\caption{\label{tab:Magnetic-Moments-of}Magnetic moments of the decuplet in
the self-consistent $\chi$QSM for $M=420$ MeV. All numbers are given
with inclusion of flavor-$SU(3)$ symmetry breaking effects. The flavor-$SU(3)$
symmetric value of this work is given by $\mu_{B^{10}}=2.47\,\, Q_{10}\,\mu_{N}$.
The $\chi$QSM $\Omega^{-}$ magnetic moment agrees well with the
experimental value given by the Particle Data Group of $\mu_{\Omega^{-}}=(-2.02\pm0.05)\,\mu_{N}$
\cite{PDG:2006}. The mass factor of Eq.(\ref{eq:G_M cqsm}) is $M_{N}^{\chi QSM}=939\cdot1.36$
MeV as described in the text.}

\end{table}

Magnetic moments for the nucleon, the $N-\Delta$ and $\Delta^{+}$
are discussed in more detail in Tab.\ref{tab:Mm N ND D+}. Since the
$\chi$QSM uses the large $N_{c}$ approximation, to some extent the
large $N_{c}$ relations of \cite{LargeNc_Mag_Mom} should be fulfilled.
The relations given in that paper are exact up to the order $\mathcal{O}(N_{c}^{-2})$.
In the present approach of the $\chi$QSM there are two reasons why
this relations should not be exactly fulfilled. First, in order to
achieve numerical values the transition back to $N_{c}=3$ is done.
Second, not all $N_{c}^{-1}$ corrections are taken into account,
e.g. corrections from the translational zero-mode are not considered.
Generally, also for other decuplet magnetic moments in the $\chi$QSM
of Tab.\ref{tab:Magnetic-Moments-of} the large $N_{C}$ relations
of \cite{LargeNc_Mag_Mom} \begin{eqnarray}
\mu_{\Delta^{++}}-\mu_{\Delta^{-}} & = & \frac{9}{5}(\mu_{p}-\mu_{n})+\mathcal{O}(N_{c}^{-2})\,\,\,,\\
\mu_{\Delta^{+}}-\mu_{\Delta^{0}} & = & \frac{3}{5}(\mu_{p}-\mu_{n})+\mathcal{O}(N_{c}^{-2})\,\,\,,\\
\mu_{\Sigma_{10}^{+}}-\mu_{\Sigma_{10}^{-}} & = & \frac{3}{2}(\mu_{\Sigma^{+}}-\mu_{\Sigma^{-}})+\mathcal{O}(N_{c}^{-2})\,\,\,,\\
\mu_{\Xi_{10}^{0}}-\mu_{\Xi_{10}^{-}} & = & -3(\mu_{\Xi^{0}}-\mu_{\Xi^{-}})+\mathcal{O}(N_{c}^{-2})\,\,\,,\end{eqnarray}
 are satisfied up to $7$\%.

\begin{table}
\begin{tabular}{c|ccc|c|c}
\hline 
$\mu[\mu_{N}]$  & $\Omega^{0}$  & $\Omega^{0+1}$  & $\Omega^{0+1}+\delta m_{s}^{1}$  & large $N_{c}$ rel.  & exp.\tabularnewline
\hline 
$\mu_{p}$  & $1.25$  & $2.46$  & $2.44$  &  & $2.79$\tabularnewline
$\mu_{n}$  & $-0.93$  & $-1.63$  & $-1.68$  &  & $-1.91$\tabularnewline
$|\mu_{\Delta N}|$  & $1.38$  & $2.56$  & $2.72$  & $\mu_{\Delta N}=\frac{1}{\sqrt{2}}(\mu_{p}-\mu_{n})=2.91$  & $3.46\pm0.03$\tabularnewline
$\mu_{\Delta^{+}}$  & $1.16$  & $2.47$  & $2.35$  & $\mu_{\Delta^{+}}\approx\frac{3}{5}(\mu_{p}-\mu_{n})=2.47$  & $2.7\pm1.15^{(stat)}\pm1.5^{(syst)}$\tabularnewline
\hline
\end{tabular}

\caption{\label{tab:Mm N ND D+}Magnetic moments of the nucleon, the $N$-$\Delta$
transition and the $\Delta^{+}$ in the self-consistent $\chi$QSM
for $M=420$MeV. The second column corresponds to the leading order
in rotation whereas the third and forth columns are linear rotational
and $m_{s}$ corrections, respectively. The last column are experimental
data taken from \cite{PDG:2006,Tiator:NDMagMom,Kotulla:Delta_mag_mom_2002,Tiator:NDquadrupole}
with the uncertainty of $\mu_{\Delta^{+}}=(2.7_{-1.3}^{+1.0}(stat.)\pm1.5(syst.)\pm3(theo.))\,\mu_{N}$.
The normalization in Eq.(\ref{eq:G_M cqsm}) is taken as $M_{N}^{\chi QSM}=939\cdot1.36$~MeV
for all given observables as described in the text. The values for
the large $N_{c}$ relations are given by using the $\chi$QSM values
where in case of the $\mu_{\Delta^{+}}$ the $\mu_{\Delta^{0}}$ contribution
is omitted.}

\end{table}

\begin{figure}
\includegraphics[scale=0.6]{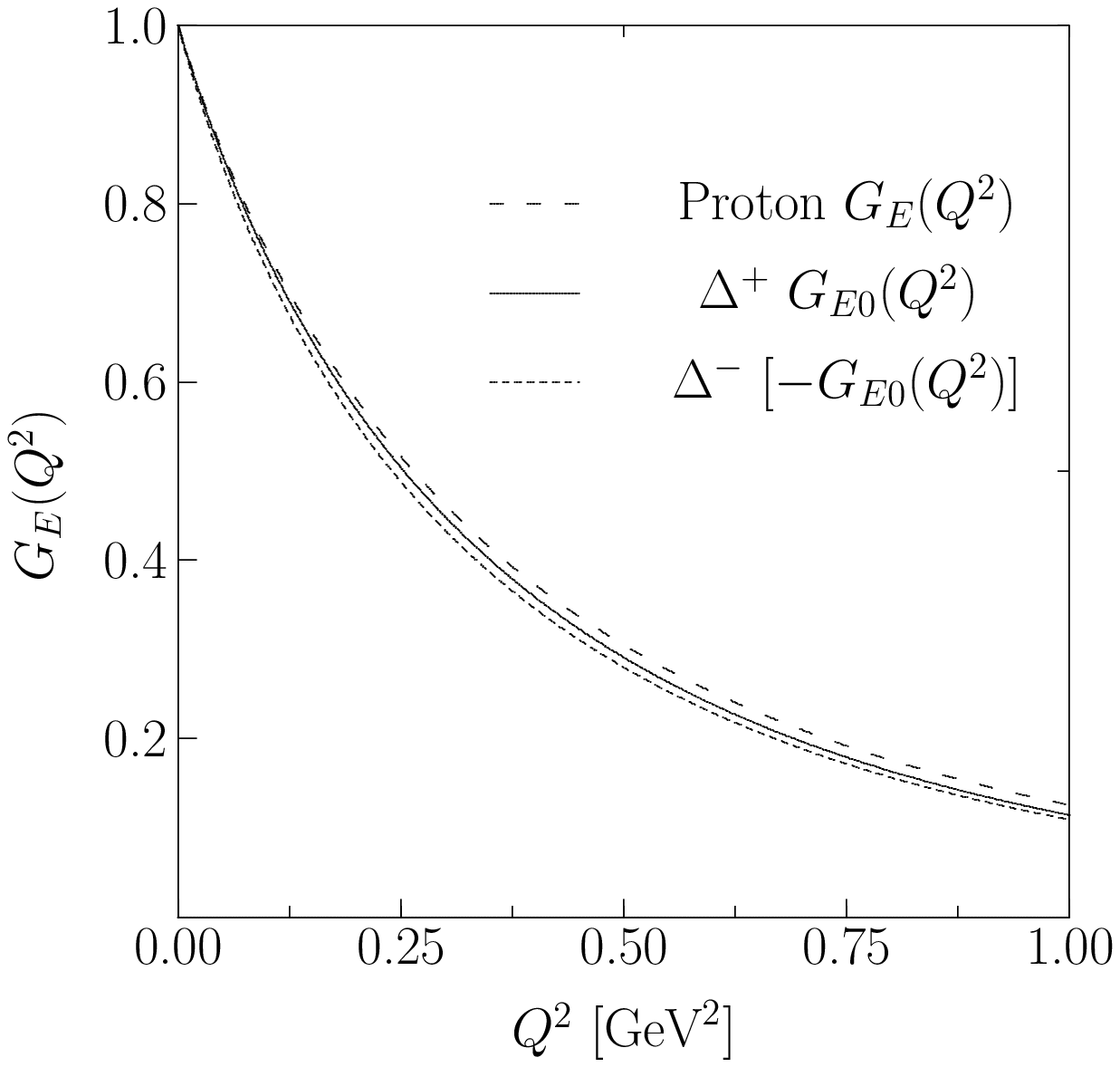}~~\includegraphics[scale=0.6]{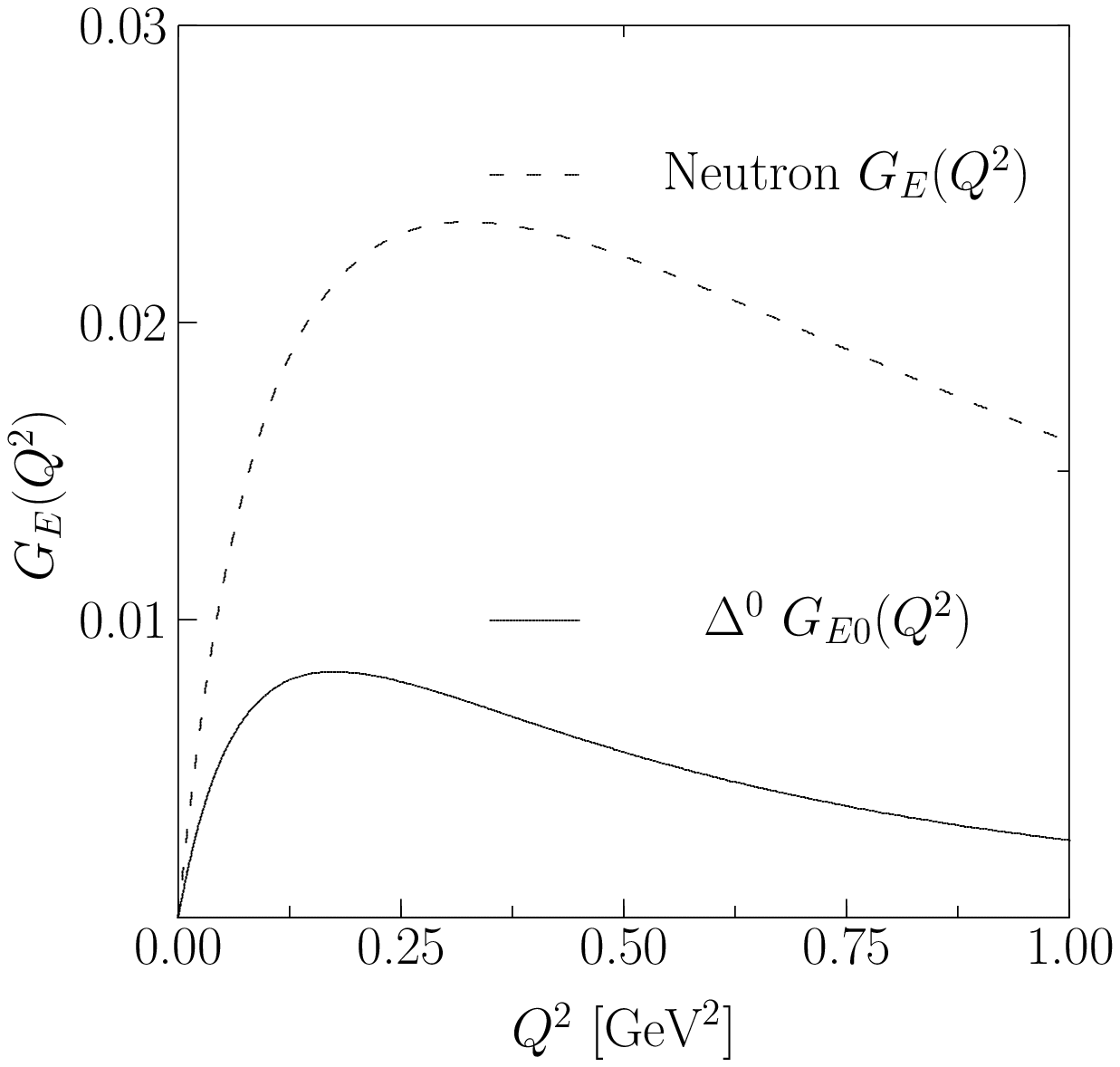}\\
 \includegraphics[scale=0.6]{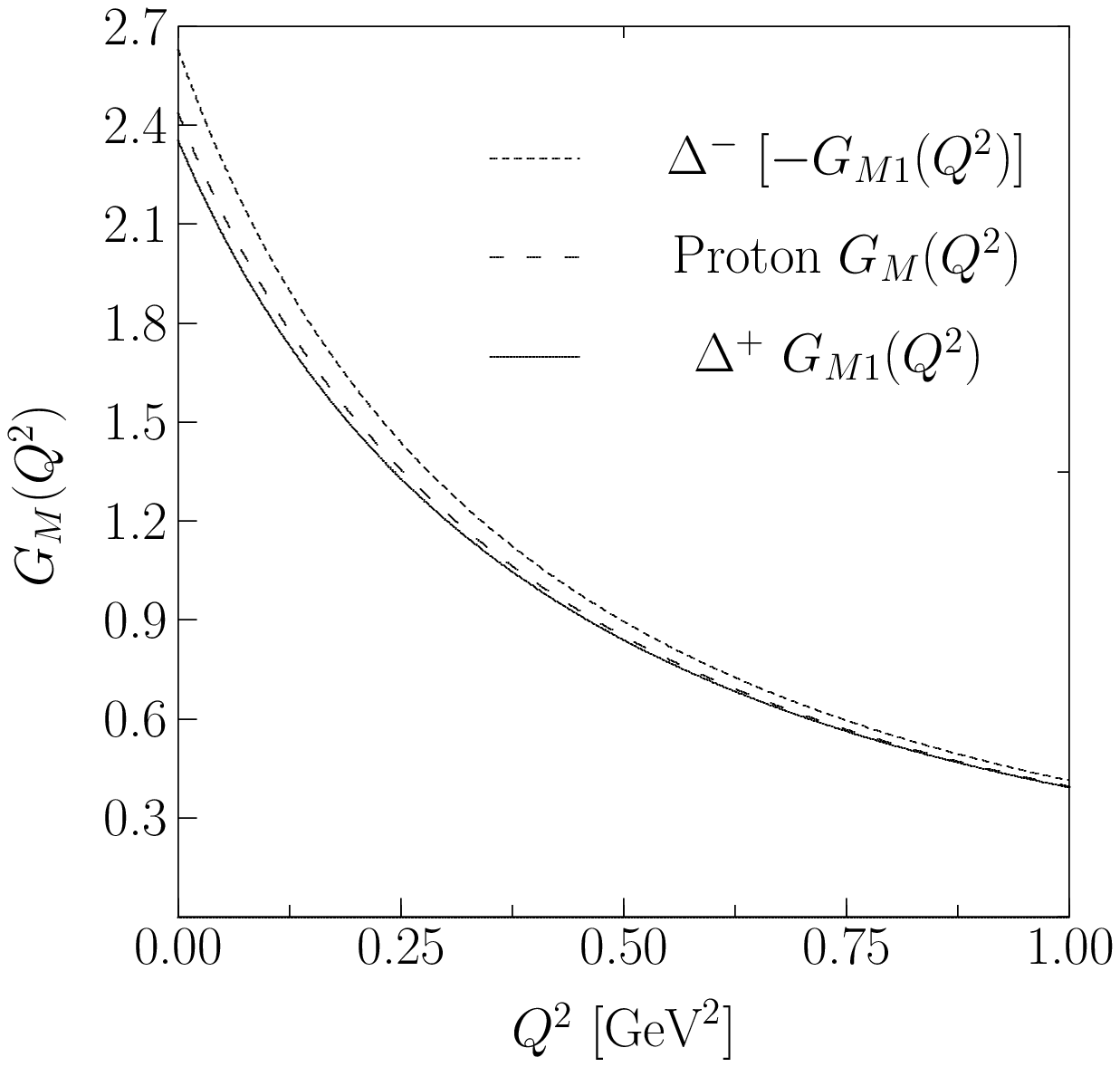}~~\includegraphics[scale=0.6]{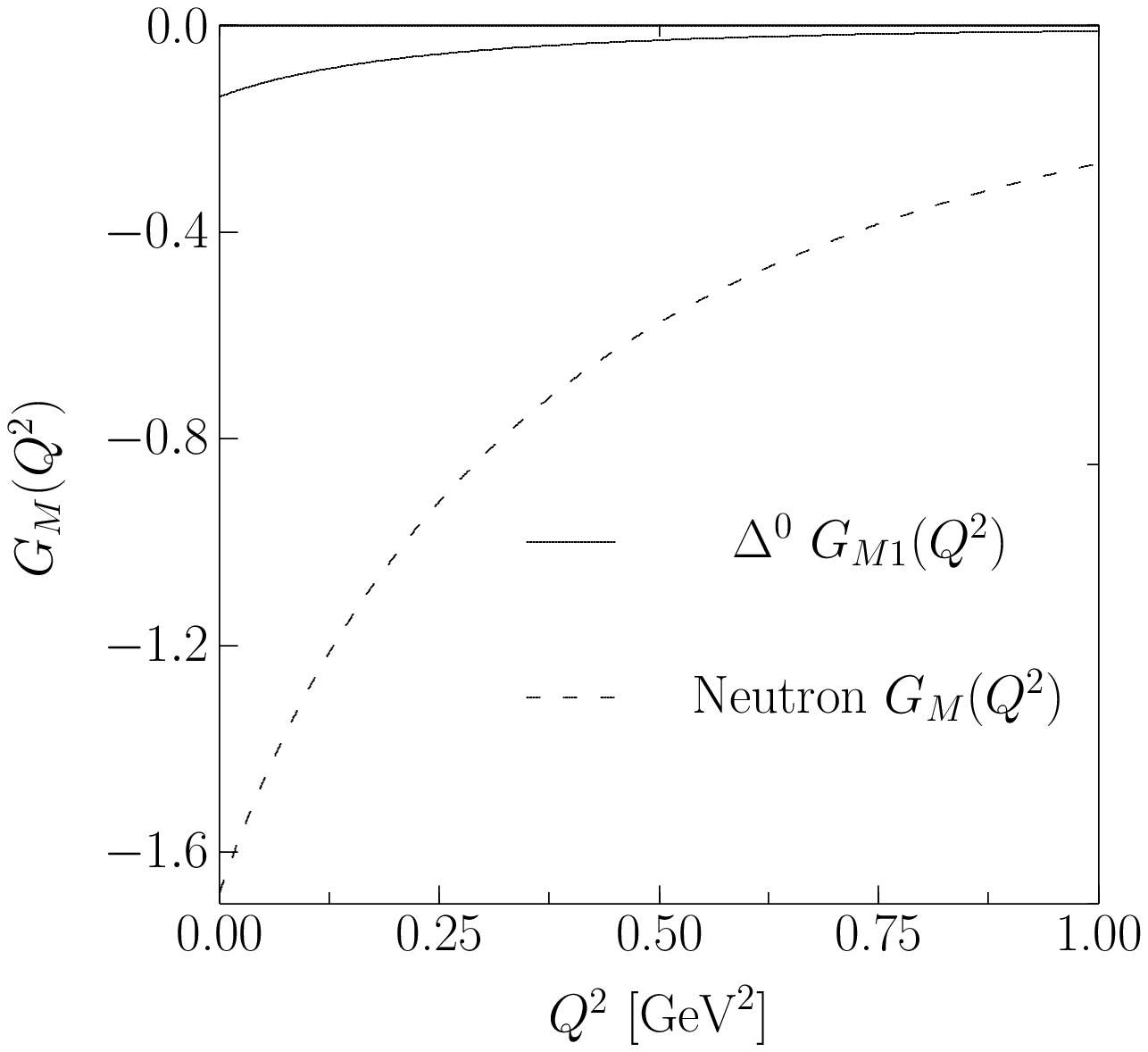}

\includegraphics[scale=0.6]{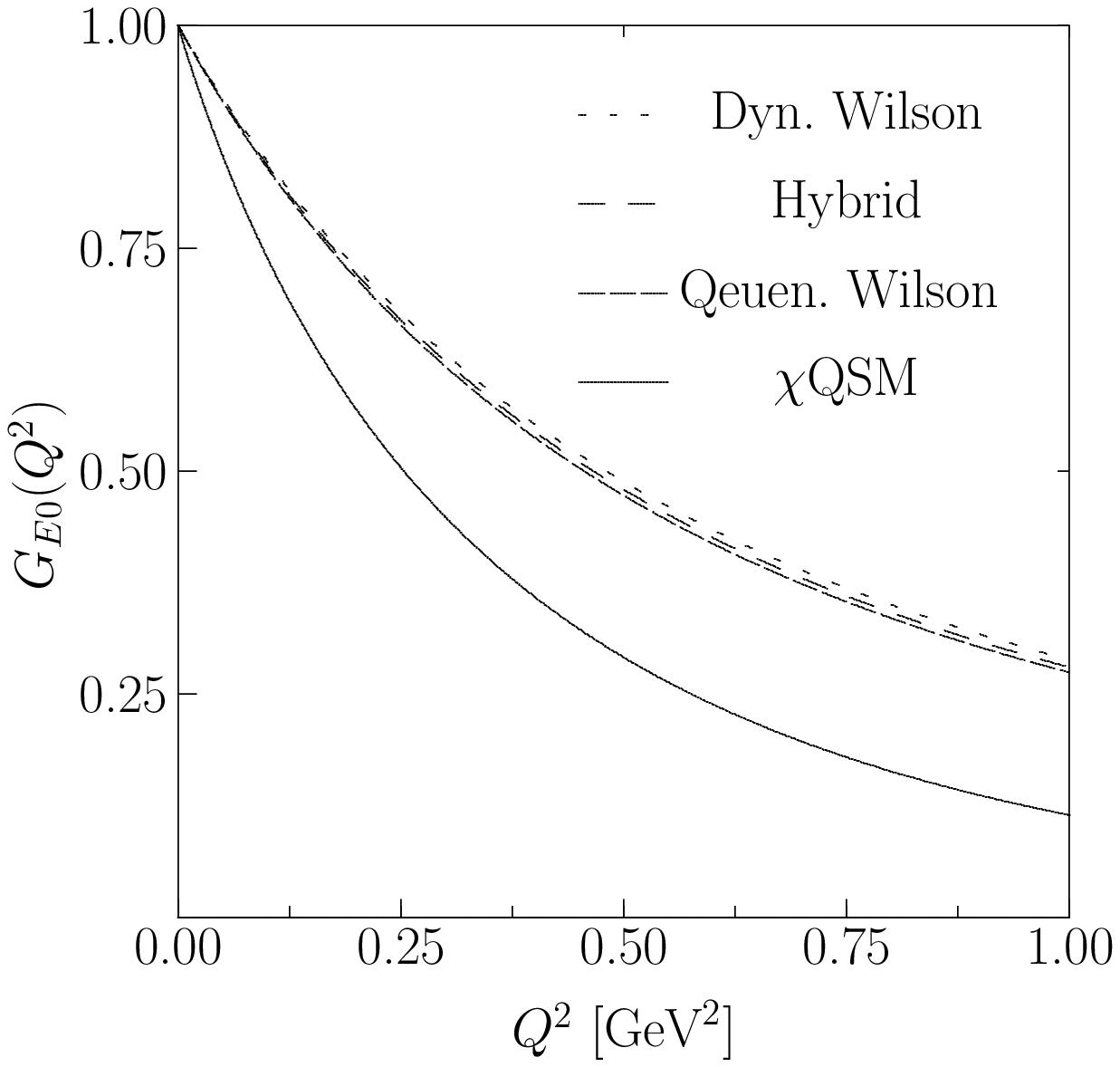}~~\includegraphics[scale=0.6]{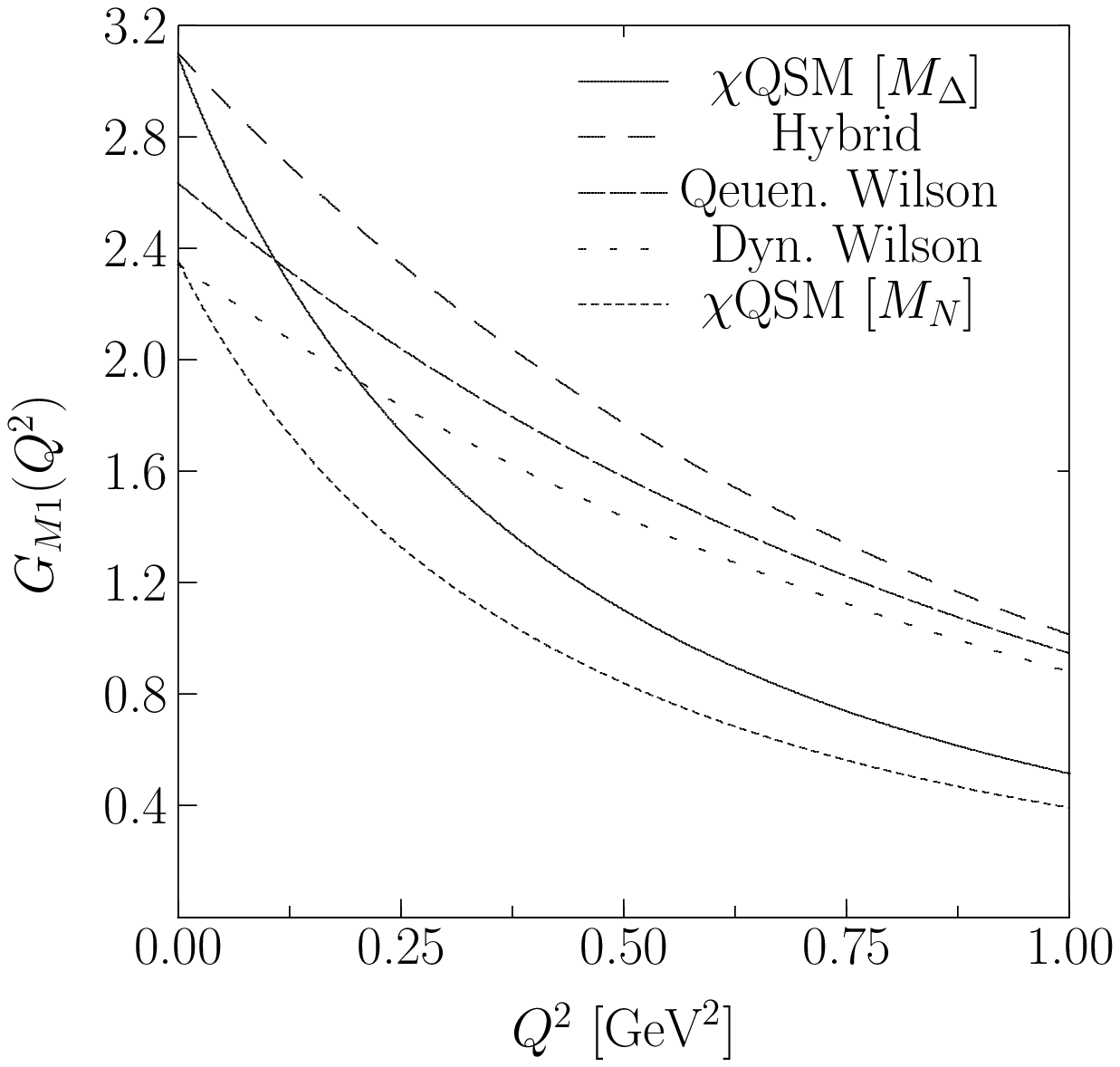}

\caption{\label{fig:El-and-ma FF}Electric and magnetic form factors of the
$\Delta^{+}$, $\Delta^{0}$, $\Delta^{-}$ and the nucleon in the
self-consistent $\chi$QSM compared to lattice results. The form factors
of the $\Delta^{++}$ are roughly by an overall factor of $2$ larger
than those for the $\Delta^{+}$ and are not explicitly shown. The
$G_{E0}^{\Delta^{0}}$ and $G_{M1}^{\Delta^{0}}$ for $Q^{2}>0$ are
entirely due to $m_{s}$ corrections and therefore smaller compared
to the neutron $G_{M}$. For all magnetic form factors $M_{N}^{\chi QSM}=939\cdot1.36$
MeV is used in Eq.(\ref{eq:G_M cqsm}) beside the $\chi$QSM graph
in the lower-right picture where we also take $M_{\Delta}^{\chi QSM}=1232\cdot1.36$
MeV and indicate the normalization by $[M_{N(\Delta)}]$. In the last
two figures we compare our final results for the $\Delta^{+}$ form
factors $G_{E0}$ and $G_{M1}$ with those of the lattice results
in \cite{Lattice_Delta_2008}. }

\end{figure}

In case of the $N-\Delta$ transition and the $\Delta$ form factors
we made use of large $N_{c}$ arguments in Eqs.(\ref{eq:largeNC1},\ref{eq:largeNC2})
for several mass-ratio factors, which lead to the values, also presented
in the Tab.\ref{tab:Mm N ND D+} and Tab.\ref{tab: fit-Para}, in
the self-consistent $\chi$QSM of

\begin{eqnarray}
G_{M}^{N\Delta}(0)=2.72 & \,\,\,\,\,\,\,\,\,\,\,\,\,\,\,\,\,\,\,\, & \mu_{N\Delta}=2.72\,\,\mu_{N}\,\,\,,\\
G_{M1}^{\Delta^{+}}(0)=2.35 &  & \mu_{\Delta^{+}}=2.35\,\,\mu_{N}\,\,\,.\end{eqnarray}
 Keeping these mass-ratio factors, which are over-all factors, yields\begin{eqnarray}
G_{M}^{N\Delta}(0)=2.30 & \,\,\,\,\,\,\,\,\,\,\,\,\,\,\,\,\,\,\,\, & \mu_{N\Delta}=2.72\,\,\mu_{N}\,\,\,,\\
G_{M1}^{\Delta^{+}}(0)=3.09 &  & \mu_{\Delta^{+}}=2.35\,\,\mu_{N}\,\,\,.\end{eqnarray}
 The first treatment would correspond to neglecting \emph{all} $1/N_{c}$
corrections beside the rotational corrections while keeping the pre-factors
would correspond to keeping some more $1/N_{c}$ corrections but neglecting
all model-based $1/N_{c}$ corrections besides the rotational ones.
\\
 We will discuss now the $\Delta^{+}$ electric and magnetic form
factors $G_{E0}$ and $G_{M1}$ for $Q^{2}\leq1\,\mbox{GeV}^{2}$.\\
 The results of the self-consistent $\chi$QSM calculations for the
electric and magnetic form factors $G_{E}^{p}$, $G_{M}^{p}$, $G_{E0}^{\Delta^{+}}$
and $G_{M1}^{\Delta^{+}}$ are best reproduced by a dipole type form
factor

\begin{equation}
G_{E,M}(Q^{2})=\frac{G_{E,M}(0)}{\Big(1+\frac{Q^{2}}{\Lambda_{E,M}^{2}}\Big)^{2}}\,\,\,.\label{eq:dipole fit}\end{equation}
 In Tab.\ref{tab: fit-Para} we present the fitted parameter which
reproduce the proton and $\Delta^{+}$ electric and magnetic form
factors of Fig.\ref{fig:El-and-ma FF}. In case of the lattice results
\cite{Lattice_Delta_2008} an exponential type form factor for $G_{M1}$
\begin{equation}
G_{M1}(Q^{2})=G_{M1}(0)\,\, e^{-Q^{2}/\Lambda_{M1}^{2}}\,\,\,,\label{eq:exp fit}\end{equation}
 parametrizes best the lattice results. We compare our results in
Tab.\ref{tab: fit-Para} with those of \cite{Lattice_Delta_2008}.

\begin{table}
\begin{tabular}{c|ccc|ccc}
\hline 
 & $\Lambda_{E}^{2}/(\mbox{GeV}^{2})$  & $G_{M}^{p}(0)$  & $\Lambda_{M}^{2}/(\mbox{GeV}^{2})$  & $\Lambda_{E0}^{2}/(\mbox{GeV}^{2})$  & $G_{M1}^{\Delta^{+}}(0)$  & $\Lambda_{M1}^{2}/(\mbox{GeV}^{2})$\tabularnewline
\hline 
$\chi$QSM  & $0.614$  & $2.438$  & $0.716$  & $0.585$  & $2.354\,\,[3.089]^{1)}$  & $0.736^{\mbox{dip}}\,\,[0.490]^{\mbox{exp}}$\tabularnewline
\hline 
Quenched Wilson  & $ $  & $ $  & $ $  & $1.101$  & $2.635$  & $0.978^{\mbox{exp}}$\tabularnewline
Dyn. $N_{f}=2$ Wilson  & $ $  & $ $  & $ $  & $1.161$  & $2.344$  & $1.022^{\mbox{exp}}$\tabularnewline
Hybrid  & $ $  & $ $  & $ $  & $1.126$  & $3.101$  & $0.895^{\mbox{exp}}$\tabularnewline
\hline 
Experiment  & $0.523$  & $2.793$  & $ $  & $ $  & $ $  & $ $\tabularnewline
\hline
\end{tabular}

\caption{\label{tab: fit-Para}Table of proton and $\Delta^{+}$ parameters
for the dipole (dip) and exponential (exp) form factor fits Eqs.(\ref{eq:dipole fit},\ref{eq:exp fit}).
The numbers in parentheses corresponds to 1) using normalizaion of
$M_{\Delta}^{\chi QSM}=1232\cdot1.36$ MeV in Eq.(\ref{eq:G_M1 cqsm})
2) using an exponential type form factor. The self-consistent $\chi$QSM
calculation for $G_{M1}$ is best reproduced by a dipole type form
factor while the numbers for $\Lambda_{M1}^{2}$ in case of the lattice
results are for an exponential type form factor.}

\end{table}

The charge and magnetic dipole form factors of the decuplet baryons
in case of flavor-$SU(3)$ symmetry can be written as

\begin{eqnarray*}
G_{E0}(Q^{2}) & = & Q_{10}\times\int d^{3}z\, j_{0}(|\vec{q}||\vec{z}|)\Big[\frac{1}{24}\mathcal{B}(\vec{z})+\frac{5}{8}\frac{\mathcal{I}_{1}(\vec{z})}{I_{1}}+\frac{1}{4}\frac{\mathcal{I}_{2}(\vec{z})}{I_{2}}\Big]\,\,\,,\\
G_{M1}(Q^{2}) & = & Q_{10}\times J_{3}\times\frac{M_{\Delta}}{12}\int d^{3}z\frac{j_{1}(|\vec{q}||\vec{z}|)}{|\vec{q}||\vec{z}|}\Big[\sqrt{3}\mathcal{Q}_{0}(\vec{z})-\frac{1}{2}\frac{\mathcal{X}_{1}(\vec{z})}{I_{1}}+\frac{\sqrt{3}}{2}\frac{\mathcal{X}_{2}(\vec{z})}{I_{2}}-\sqrt{\frac{1}{2}}\frac{\mathcal{Q}_{1}(\vec{z})}{I_{1}}\Big]\,\,\,,\end{eqnarray*}
 with $Q_{10}$ as the charge of the decuplet baryon and its third-spin
component $J_{3}$ and $M_{\Delta}$ the normalization of the magnetic
form factor. In case of the neutral decuplet baryons the entire form
factors for $G_{E0}$ and $G_{M1}$, even for $Q^{2}>0$, are only
due to strange-quark mass corrections.\\
 For the proton the experimental value of the charge radius is $[\langle r_{E}^{2}\rangle^{P}]^{1/2}=0.8750\pm0.0068\,\mbox{fm}$
($\langle r_{E}^{2}\rangle^{P}\approx0.766\,\mbox{fm}^{2}$) \cite{PDG:2006}.\\
 The charge radii of the proton and $\Delta^{+}$ of $G_{E}$ and
$G_{E0}$ in the self-consistent $\chi$QSM with $M=420$ MeV are,
respectively\begin{eqnarray}
\langle r_{E}^{2}\rangle_{P} & = & 0.768\,\mbox{fm}^{2}\,\,\,\,\,\,\,\,\,\,\,\:\langle r_{E}^{2}\rangle_{P}^{SU(3)}=0.770\,\mbox{fm}^{2}\,\,\,,\\
\langle r_{E}^{2}\rangle_{\Delta^{+}} & = & 0.794\,\mbox{fm}^{2}\,\,\,\,\,\,\,\,\,\,\,\:\langle r_{E}^{2}\rangle_{\Delta^{+}}^{SU(3)}=0.813\,\mbox{fm}^{2}\,\,\,,\end{eqnarray}
 and the magnetic radii for $G_{M}(Q^{2})$ and $G_{M1}(Q^{2})$ are\begin{eqnarray}
\langle r_{M}^{2}\rangle_{P} & = & 0.656\,\mbox{fm}^{2}\,\,\,\,\,\,\,\,\,\,\,\:\langle r_{M}^{2}\rangle_{P}^{SU(3)}=0.665\,\mbox{fm}^{2}\,\,\,,\\
\langle r_{M}^{2}\rangle_{\Delta^{+}} & = & 0.634\,\mbox{fm}^{2}\,\,\,\,\,\,\,\,\,\,\,\:\langle r_{M}^{2}\rangle_{\Delta^{+}}^{SU(3)}=0.658\,\mbox{fm}^{2}\,\,\,,\end{eqnarray}
 where the index $SU(3)$ indicates the value in case of flavor-$SU(3)$
symmetry. The above radii are calculated by differentiating the $\chi$QSM
form factor expression, i.e. explicitly integrating the $\chi$QSM
form factor densities. Alternatively one could calculate the radii
by using the dipole fit due to $\langle r_{E,M}^{2}\rangle=12/\Lambda_{E,M}^{2}$
for which the values only differ by max 1\%.\\
 In Fig.\ref{fig:El-and-ma FF} we compare the final $\chi$QSM results
for the $\Delta^{+}$ form factors $G_{E0}$ and $G_{M1}$ with those
of the lattice calculation \cite{Lattice_Delta_2008}. The $\chi$QSM
form factors drop faster with increasing $Q^{2}$. In case of the
$\chi$QSM it is known that the $Q^{2}$ dependence of the experimental
data of the electric and magnetic form factors for both nucleons are
very well reproduced \cite{Christov:1995vm}. In the lattice work
\cite{Lattice_N_2006} the nucleon iso-vector form factor $F_{1}^{p-n}(Q^{2})$
for pion-masses ranging from $m_{\pi}=775$ MeV down to $m_{\pi}=359$
MeV was calculated. It was found that the form factor becomes steeper
by lowering the pion-mass. Still for a value of $m_{\pi}=359$ MeV
the results of \cite{Lattice_N_2006} are above the experimental
values. The minimal value of $m_{\pi}$ in Ref.\cite{Lattice_Delta_2008}
for the form factors $G_{E0}$ and $G_{M1}$ of the $\Delta^{+}$,
Fig.\ref{fig:El-and-ma FF}, is $m_{\pi}=353$ MeV and also do not
fall off as fast as the $\chi$QSM results. This can also be seen
in the fact, that the lattice results are best reproduced by an exponential
type form factor while the $\chi$QSM are more of a dipole type form
factor. The $\Delta$ magnetic moment is presented in the range of
$\mu_{\Delta^{+}}=(1.58\sim1.91)\,\mu_{N}$ in the pion mass range
$m_{\pi}\approx(353-400)$ MeV. The value of the present $\chi$QSM
calculation is $\mu_{\Delta^{+}}=2.35\,\mu_{N}$.\\
Recently, a first dynamical lattice QCD calculation \cite{Aubin:FiniteVstudy}
of the $\Delta$ and $\Omega^{-}$ magnetic dipole moments was also
performed using a background field method. The calculation for $\Omega^{-}$
was done at the physical strange quark mass, with the result $\mu_{\Omega^{-}}=-1.93(8)\,\mu_{N}$
in very good agreement with the experimental number. The $\Delta$
has been studied at smallest pion mass value $m_{\pi}=366$ MeV with
the result $\mu_{\Delta^{+}}=2.40(6)\,\mu_{N}$. 

We will now discuss the results for the $\Delta^{+}$ electric quadrupole
form factor $G_{E2}$. In Fig.\ref{fig:G_{E2}(Q^{2})} we present
the final results and compare them with the recent lattice calculations
in \cite{Lattice_Delta_2008}. As already mentioned in Sec.\ref{sub:DeltaDeltaCQSM}
the form factor $G_{E2}$ in the $\chi$QSM is only due to rotational
corrections which are seen as $1/N_{c}$ corrections. In the large
$N_{c}$ limit the $\chi$QSM leads to a vanishing form factor. In
the left panel of Fig.\ref{fig:G_{E2}(Q^{2})} we decomposed the form
factor into its contributions coming from the valence and sea quarks.
The sea contribution gives the most sizeable part of the form factor.
This behavior is also seen in Ref.\cite{Watabe:ND} where the electric
quadrupole moment $Q_{N\Delta}$ was investigated in the $SU(2)$
$\chi$QSM. The density $\mathcal{I}_{1E2}(r)$ also contributes to
the $N\Delta$ transition in \cite{Watabe:ND}. The Fig.\ref{fig:G_{E2}(Q^{2})}
shows the same behavior of valence and sea quark contributions for
$G_{E2}$ as Fig.1 in Ref.\cite{Watabe:ND} for the quantity $Q_{N\Delta}$.
In case of the $\chi$QSM we had to introduce a regularization scheme
for the sea quark contribution which was the proper-time regularization.
The fact that the sea quarks give the dominant part of the form factor
could result in a sensibility of the $\chi$QSM $G_{E2}$ to the applied
regularization scheme. An analogous situation is met, and well known,
in case of the $\Sigma_{\pi N}$ form factor in \cite{Blotz:SigmaPiN,Kim:SigmaPiN}.
In this work we do not investigate the regularization dependence of
the form factor $G_{E2}$ and give all final results for applying
the proper-time regularization.\\
For the parametrization of this form factor we prefer a dipole type
fit Eq.(\ref{eq:dipole fit}). In Tab.\ref{tab:fit GE2} we summarize
the parameters which reproduce the self-consistent $\chi$QSM calculation
and compare them to the results of the lattice calculation of \cite{Lattice_Delta_2008}.
In case of the electric quadrupole form factor the lattice results
are more divergent. Again the $\chi$QSM result falls off faster in
the region $0\leq Q^{2}\leq0.50\,\mbox{GeV}^{2}$ compared to all
three lattice results but compares well to the quenched Wilson and
hybrid action results for $0.50\,\mbox{GeV}^{2}\leq Q^{2}\leq1\,\mbox{GeV}^{2}$,
respectively.

\begin{figure}
\includegraphics[scale=0.6]{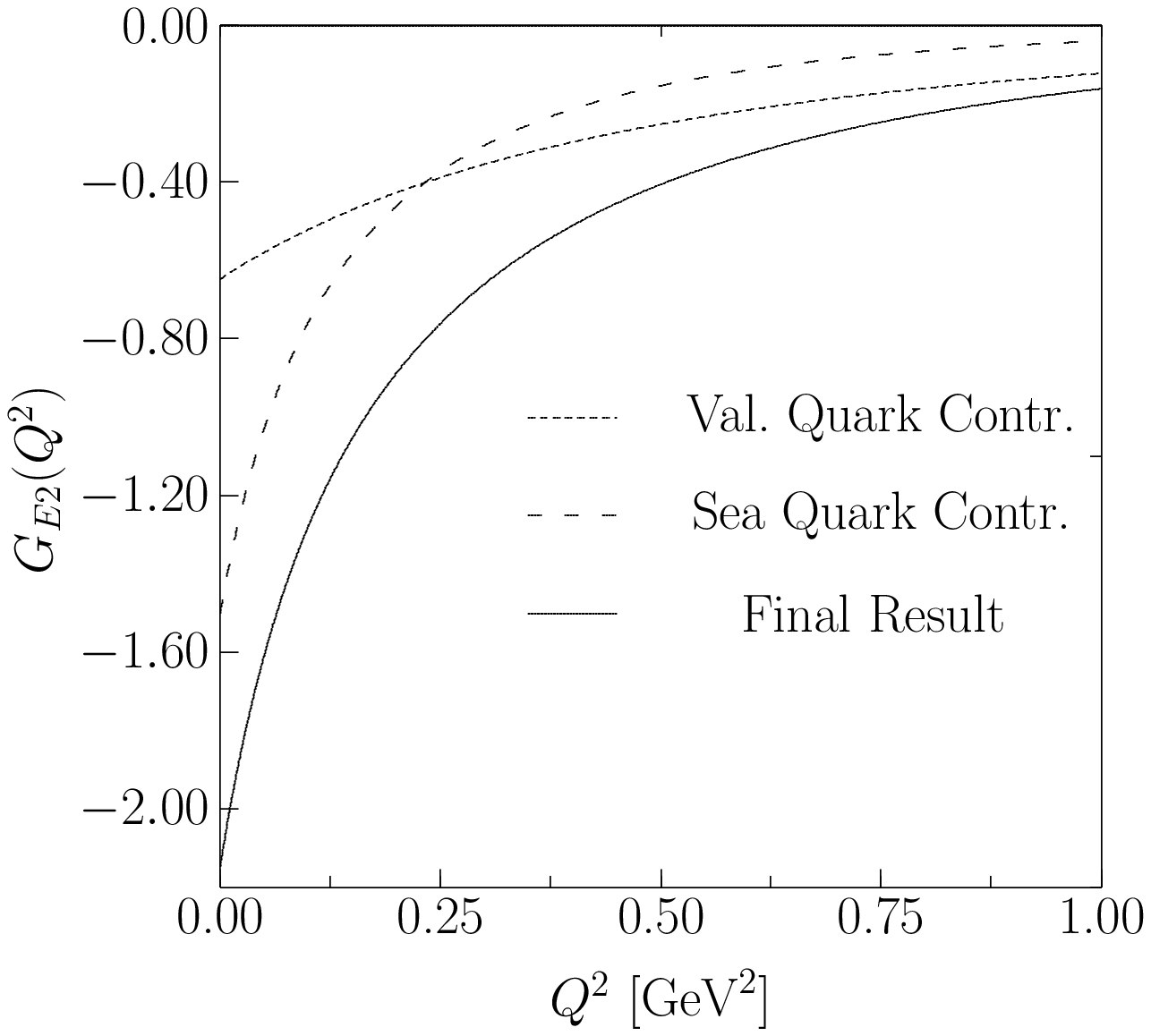}~~\includegraphics[scale=0.6]{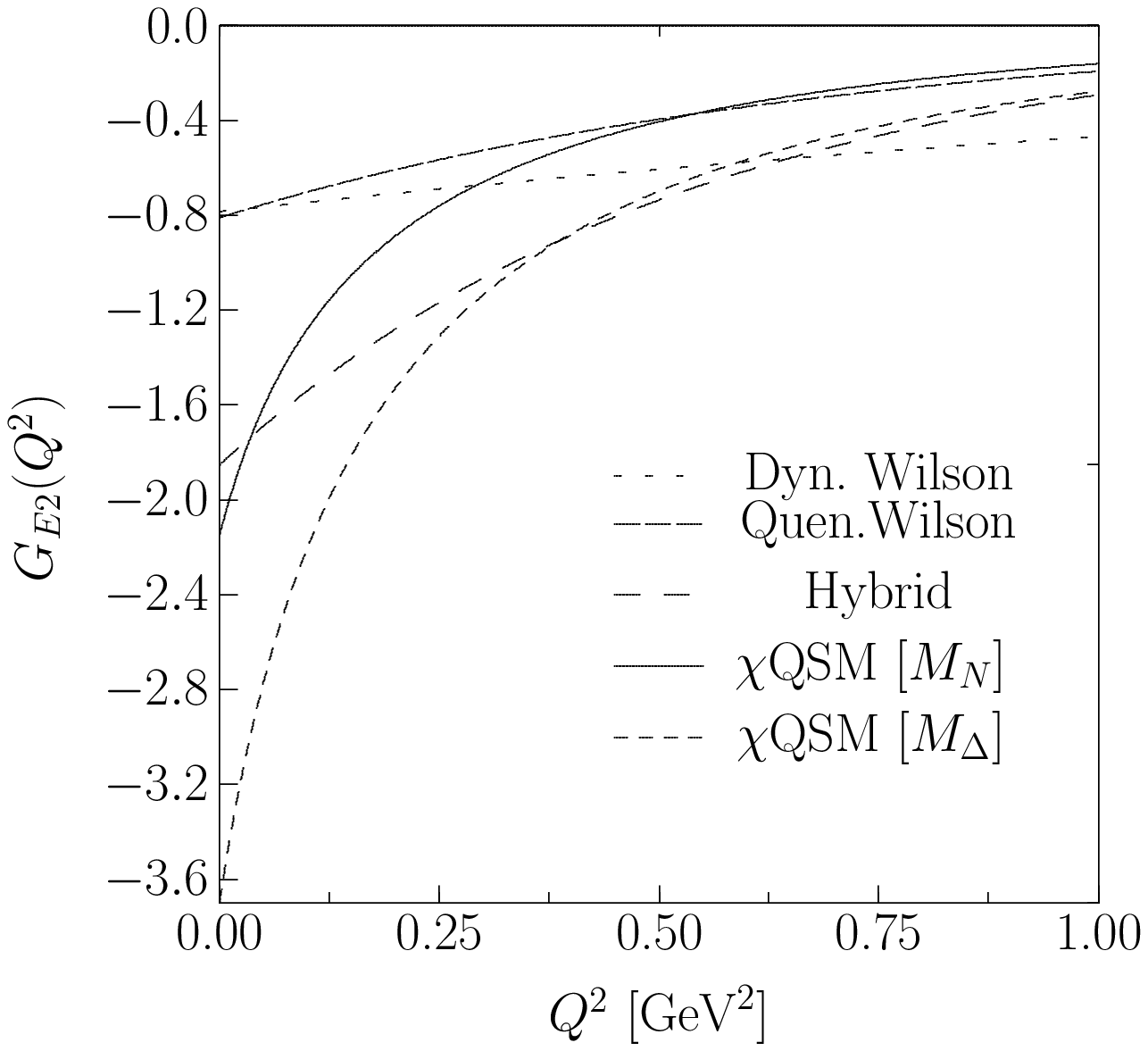}

\caption{\label{fig:G_{E2}(Q^{2})}The electric $\Delta^{+}$ quadrupole form
factor $G_{E2}$ in the self-consistent $\chi$QSM and comparison
to the lattice results of \cite{Lattice_Delta_2008}. The left picture
shows the form factor decomposed into its valence and sea quark contributions
while the right picture compares the final result with those of the
lattice calculation. In the right picture we once took $M_{N}^{\chi QSM}=939\cdot1.36$MeV
and once $M_{\Delta}^{\chi QSM}=1232\cdot1.36$MeV for the mass in
Eq.(\ref{eq:GE2 CQSM Final}). }

\end{figure}

\begin{table}
\begin{tabular}{c|c|ccc}
\hline 
 & $\chi$QSM  & Quenched Wilson  & Dynamical Wilson  & Hybrid\tabularnewline
\hline 
$G_{E2}^{\Delta^{+}}(0)$  & $-2.145$  & $-0.810$  & $-0.784$  & $-1.851$\tabularnewline
$\Lambda_{E2}^{2}/(\mbox{GeV}^{2})$  & $0.369{}^{\mbox{dip}}\,\,[0.268]^{\mbox{exp}}$  & $0.696^{\mbox{exp}}$  & $1.938^{\mbox{exp}}$  & $0.542^{\mbox{exp}}$\tabularnewline
\hline
\end{tabular}

\caption{\label{tab:fit GE2}Table for fit parameters of the form factor $G_{E2}$.
The indices {}``dip'' and {}``exp'' corresponds to fitting with
a dipole or exponential type form factor Eqs.(\ref{eq:dipole fit},\ref{eq:exp fit}).
A dipole type form factor reproduces the self-consistent $\chi$QSM
calculation more accurate than an exponential fit. }

\end{table}

In the Ref.\cite{Large_NC_quadrupole} a relation in the large $N_{c}$
limit is found which connects the quadrupole moment of the $N-\Delta$
transition $Q_{N\Delta}$ to the quadrupole moment $Q_{\Delta}$ of
the $\Delta$

\begin{equation}
Q_{\Delta^{+}}=\frac{2\sqrt{2}}{5}Q_{p\Delta^{+}}+\mathcal{O}(N_{c}^{-2})\,\,\,.\end{equation}
 The Ref.\cite{Tiator:NDquadrupole} extracted the value of \begin{equation}
Q_{N\Delta}=-(0.0846\pm0.0033)\,\mbox{fm}^{2}\,\,\,,\end{equation}
 which gives with the above large $N_{c}$ relation \[
Q_{\Delta^{+}}=(-0.048\pm0.002)\,\mbox{fm}^{2}\,\,\,.\]
 The final result of this work in the self-consistent $\chi$QSM is
\begin{equation}
Q_{\Delta^{+}}^{\chi QSM}=\frac{G_{E2}(0)}{M_{\Delta}^{2}}=-0.0509\,\mbox{fm}^{2}\,\,\,,\label{eq:QD CQSM final}\end{equation}
which agrees well to the above estimation. From the left panel in
Fig.\ref{fig:G_{E2}(Q^{2})} we see that for the electric quadrupole
moment, propotional to $G_{E2}(0)$, the sea quark contribution dominates
the valence quark contribution. Furthermore, one can expect the sea
quark contribution to have a broader spatial distribution than the
one for the valence quarks. This in turn leads to a steeper $Q^{2}$
dependence of the contribution to $G_{E2}$ of sea quarks as compared
with valence quarks. This is evidenced in the present calculation
as shown in Fig.\ref{fig:G_{E2}(Q^{2})}. \\
In the $\chi$QSM work \cite{Watabe:ND} the authors presented an
electric quadrupole transition moment of $Q_{N\Delta}=-0.020\,\mbox{fm}^{2}$.
Also for this quantity the main contribution comes from the sea quarks.
The small value of $Q_{N\Delta}=-0.020\,\mbox{fm}^{2}$ in \cite{Watabe:ND}
is in contrast to $Q_{N\Delta}=-(0.0846\pm0.0033)\,\mbox{fm}^{2}$
from \cite{Tiator:NDMagMom} and the relative large electric $\Delta$
quadrupole moment $Q_{\Delta^{+}}^{\chi QSM}=-0.0509\,\mbox{fm}^{2}$
of this work. We can reproduce with the density $\mathcal{I}_{1E2}(r)$
of this work the values given in \cite{Watabe:ND}. The discrepancy
of the above numbers could be due to a possible breakdown of the approximation
$k\cdot R\ll1$ performed in \cite{Watabe:ND}, with $k$ being the
photon-momentum at $Q^{2}=0$ of the $\gamma^{*}N\Delta$ process
and $R$ being the nucleon charge radius. This remains to be investigated
in future studies.\\
In the work \cite{Buchman:DeltaQuad} the $\Delta^{+}$ electric
quadrupole moment is estimated to $Q_{\Delta^{+}}^{imp\,(exc)}=-0.032\,\mbox{fm}^{2}\,(-0.119\,\mbox{fm}^{2})$
by using a constituent quark model with once configuration mixings
and no exchange current and once with an exchange current but no configuration
mixing, respectively. A recent light cone QCD sum rule calculation
\cite{Azizi:DeltaQuad} obtained an electric quadrupole moment of
$Q_{\Delta^{+}}=-(5.8\pm1.45)10^{-4}\,\mbox{fm}^{2}$. Our value of
$Q_{\Delta^{+}}^{\chi QSM}=-0.0509\,\mbox{fm}^{2}$ is more comparable
to the constituent quark model results.

\section{Summary}

In the present work we investigated in the framework of the self-consistent
$SU(3)$ $\chi$QSM the electromagnetic form factors of the vector
current for the decuplet baryons. We explicitly take the symmetry
conserving quantization, linear $1/N_{c}$ rotational as well as linear
strange-quark mass corrections into account. Earlier self-consitent
$SU(3)$ $\chi$QSM results only calculated the decuplet magnetic
moments and did not apply the symmetry conserving quantization. Numerical
parameters of the model are fixed in the meson-sector as described
at the end of Sec.\ref{sec:Form-factors-in}. The only free parameter
of the $\chi$QSM for the baryon-sector is then the constituent quark
mass. All these parameters were fixed by previous studies and were
also used in the present work. No additional readjusting is done.
With these parameters, the general way to calculate observables in
the model is to determine the eigenvalues of the $\chi$QSM hamiltonian
numerically by using a self-consistent pion-field profile, the soliton.
These eigenvalues are then used for determining all observables in
the $\chi$QSM.\\
In particular we calculated the form factors $G_{E0}$, $G_{M1}$
and $G_{E2}$ for the $\Delta^{+}$ up to a momentum-transfer of $Q^{2}\leq1\mbox{GeV}^{2}$
and magnetic moments for all decuplet baryons and the $N-\Delta$
transition. In general all $\chi$QSM form factors are best reproduced
by a dipole type fit.\\
Experimental data for decuplet magnetic moments are available for
the $\Delta^{++}$ with $\mu_{\Delta^{++}}=3.7\sim7.5\mu_{N}$ \cite{PDG:2006},
the $\Delta^{+}$ with $\mu_{\Delta^{+}}=(2.7_{-1.3}^{+1.0}(\mbox{stat.})\pm1.5(\mbox{syst.})\pm3(\mbox{theor.}))\,\mu_{N}$
\cite{Kotulla:Delta_mag_mom_2002} and for the $\Omega^{-}$ with
$\mu_{\Omega^{-}}=(-2.02\pm0.05)\mu_{N}$. The present work yields
values of $\mu_{\Delta^{++}}=4.85\,\mu_{N}$, $\mu_{\Delta^{+}}=2.35\,\mu_{N}$
and $\mu_{\Omega^{-}}=-2.29\,\mu_{N}$ which is in good agreement
with the experimental ones. The $N-\Delta$ magnetic transition moment
was extracted in \cite{Tiator:NDquadrupole} as $\mu_{N\Delta}=3.46\pm0.03\,\mu_{N}$
whereas this work yields a value of $\mu_{\Delta N}=2.72\,\mu_{N}$.
Other $\chi$QSM results for decuplet magnetic moments are summarized
in Tab.\ref{tab:Magnetic-Moments-of}.\\
The final results for the magnetic dipole and electric charge form
factors are presented in Figs.\ref{fig:El-and-ma FF}. In the $\chi$QSM
the $\Delta^{+}$ radii of these form factors, $\langle r_{E}^{2}\rangle=0.794\,\mbox{fm}^{2}$
and $\langle r_{M}^{2}\rangle=0.634\,\mbox{fm}^{2}$, are comparable
to the ones of the proton, $\langle r_{E}^{2}\rangle=0.768\,\mbox{fm}^{2}$
and $\langle r_{M}^{2}\rangle=0.656\,\mbox{fm}^{2}$, keeping in mind
that we take for both baryons the same classical soliton configuration.
The experimental value for the proton electric radius is $\langle r_{E}^{2}\rangle\approx0.766\,\mbox{fm}^{2}$.\\
We also presented the electric quadrupole form factor of the $\Delta^{+}$.
The value $G_{E2}(0)$ is directly proportional to the $\Delta$ electric
quadrupole moment for which we found a value of $Q_{\Delta}=-0.0509\,\mbox{fm}^{2}$.
The electric quadrupole moment and the electric quadrupole form factor
appear in the model entirely as $1/N_{c}$ corrections arising form
the expansion in the rotation velocity of the soliton. Hence, in the
large $N_{c}$ limit the model leads to a vanishing form factor and
moment. In addition a decomposition into the valence and sea quark
contribution of the electric quadrupole form factor, Fig.\ref{fig:G_{E2}(Q^{2})},
shows that the main contribution originates from the sea quarks. Furthermore,
one can expect the sea quark contribution to have a broader spatial
distribution than the one for the valence quarks. This in turn leads
to a steeper $Q^{2}$ dependence of the contribution to $G_{E2}$
of sea quarks as compared with valence quarks which is explicitly
seen in the present calculation.\\

\subsubsection*{Acknowledgments}

The authors are very grateful to K. Goeke for helpful comments. A.S.
acknowledges support from DAAD-GRICES and PTDC/FIS/64707/2006. The
present work was partially supported by the Research Centre \char`\"{}Elementarkraefte
und Mathematische Grundlagen\char`\"{} at the Johannes Gutenberg University
Mainz.

\section{Appendix}

\subsection{Model Independent Quantities}

We use the Breit-frame in which the incoming $p$ and outgoing $p^{\prime}$
momenta are defined as\begin{equation}
p^{\prime}=(E,\frac{\vec{q}}{2})\,\,\,,\,\,\, p=(E,-\frac{\vec{q}}{2})\,\,\,,\,\,\, q=(0,\vec{q})\,\,\,,\,\,\, Q^{2}=-q^{2}=\vec{q}^{2}\,\,\,,\,\,\, q=|\vec{q}|(0,\sin\theta\cos\phi,\sin\theta\sin\phi,\cos\theta)\,\,\,\end{equation}
 with $\vec{q}^{\,2}=4(E^{2}-M^{2})$. We use the Rarita-Schwinger
spin-3/2 spinor \[
u^{\alpha}(p,s)=\sum_{\lambda,s^{\prime}}\,\, C_{1\lambda\frac{1}{2}s^{\prime}}^{\frac{3}{2}s}\,\, e^{\alpha}(p,\lambda)\,\, u(p,s^{\prime})\,\,\,\,\,\,\,\,\,\,\,\,\,\,\,\,\mbox{with}\,\,\,\,\,\,\,\,\,\,\,\,\, u(p,s)=\sqrt{\frac{E+M}{2M}}\left(\begin{array}{c}
\phi_{s}\\
\frac{\vec{\sigma}\cdot\vec{p}}{E+M}\phi_{s}\end{array}\right)\,\,\,.\]
 The spin-1 vector $e^{\alpha}(p,\lambda)$ is defined by with $\lambda=\pm1,0$

\begin{equation}
e^{\alpha}(p,\lambda)=\Big(\frac{\hat{e}_{\lambda}\cdot\vec{p}}{M},\hat{e}_{\lambda}+\frac{\vec{p}\cdot(\hat{e}_{\lambda}\cdot\vec{p})}{M(p^{0}+M)}\Big)\,\,\,\mbox{with}\,\,\,\hat{e}_{+1}=\sqrt{\frac{1}{2}}\left(\begin{array}{c}
-1\\
-i\\
0\end{array}\right)\,\,\,,\,\,\,\hat{e}_{0}=\left(\begin{array}{c}
0\\
0\\
1\end{array}\right)\,\,\,,\,\,\,\hat{e}_{-1}=\sqrt{\frac{1}{2}}\left(\begin{array}{c}
1\\
-i\\
0\end{array}\right)\,\,\,.\end{equation}
 The final and initial $\Delta$ states for the used third-spin components
read

\begin{eqnarray}
u^{\beta}(p,+\frac{3}{2}) & = & u(p,+\frac{1}{2})\,\, e^{\beta}(p,+1)\\
u^{\beta}(p,+\frac{1}{2}) & = & \sqrt{\frac{2}{3}}\,\, u(p,+\frac{1}{2})\,\, e^{\beta}(p,0)+\sqrt{\frac{1}{3}}\,\, u(p,-\frac{1}{2})\,\, e^{\beta}(p,+1)\,\,\,.\end{eqnarray}
 For the zeroth-component of the vector current, $\langle\Delta(\frac{3}{2})|V^{0}|\Delta(\frac{3}{2})\rangle$
we obtain by using the Breit-frame

\begin{eqnarray}
\overline{u}(p^{\prime},s^{\prime})\gamma^{0}u(p,s) & = & \delta_{s^{\prime}s}\,\,\,\,\,\,\,\,\,\,\,\,;\,\,\, e^{*\alpha}(p^{\prime},1)g_{\alpha\beta}e^{\beta}(p,1)=-1-\frac{2}{3}\tau+(3\cos^{2}\theta-1)\frac{\tau}{3}\,\,\,,\\
\overline{u}(p^{\prime},s^{\prime})\sigma^{0\nu}q_{\nu}u(p,s) & = & -i\frac{q^{2}}{2M}\delta_{s^{\prime}s}\,\,\,;\,\,\, e^{*\alpha}(p^{\prime},1)q_{\alpha}q_{\beta}e^{\beta}(p,1)=4M^{2}\tau\frac{1+\tau}{3}[1-\frac{1}{2}(3\cos^{2}\theta-1)]\,\,\,,\end{eqnarray}
 with $\tau=Q^{2}/(4M^{2})$.\\
 For the spatial-component of the vector current $\langle\Delta|V_{k}|N\rangle$we
obtain by using the rest-frame of the $\Delta$

\begin{eqnarray}
\epsilon_{\beta k\sigma\tau}P_{\sigma}q_{\tau} & = & M\,\,\delta^{\beta b}\,\,\epsilon^{bks}q^{s}\,\,\,,\\
\epsilon_{\beta\sigma\nu\gamma}P_{\nu}q_{\gamma}\epsilon_{k\sigma\alpha\delta}p_{\alpha}^{\prime}q_{\delta}=\epsilon_{\beta\sigma\nu\gamma}P_{\nu}q_{\gamma}\epsilon_{k\sigma0\delta}Mq_{\delta} & = & M^{2}\,\,\delta^{\beta b}\,\,[\delta^{bk}\vec{q}^{2}-q^{b}q^{k}]\,\,\,.\end{eqnarray}

\subsection{$\chi$QSM Electric Densities}

The electric densities of Eq.(\ref{eq:Ele dens}) are \begin{eqnarray*}
\frac{1}{N_{c}}\mathcal{B}(\vec{z}) & = & \phi_{v}^{\dagger}(\vec{z})\phi_{v}(\vec{z})-\frac{1}{2}\sum_{n}\textrm{sign}(\varepsilon_{n})\phi_{n}^{\dagger}(\vec{z})\phi_{n}(\vec{z}),\\
\frac{1}{N_{c}}\mathcal{I}_{1}(\vec{z}) & = & \frac{1}{2}\sum_{\varepsilon_{n}\neq\varepsilon_{v}}\frac{1}{\varepsilon_{n}-\varepsilon_{v}}\langle v|\tau^{i}|n\rangle\phi_{n}^{\dagger}(\vec{z})\tau^{i}\phi_{v}(\vec{z})+\frac{1}{4}\sum_{n,m}\mathcal{R}_{3}(\varepsilon_{n},\varepsilon_{m})\langle n|\tau^{i}|m\rangle\phi_{m}^{\dagger}(\vec{z})\tau^{i}\phi_{n}(\vec{z}),\\
\frac{1}{N_{c}}\mathcal{I}_{2}(\vec{z}) & = & \frac{1}{4}\sum_{\varepsilon_{n^{0}}}\frac{1}{\varepsilon_{n^{0}}-\varepsilon_{v}}\langle n^{0}|v\rangle\phi_{v}^{\dagger}(\vec{z})\phi_{n^{0}}(\vec{z})+\frac{1}{4}\sum_{n,m^{0}}\mathcal{R}_{3}(\varepsilon_{n},\varepsilon_{m^{0}})\phi_{m^{0}}^{\dagger}(\vec{z})\phi_{n}(\vec{z})\langle n|m^{0}\rangle,\\
\frac{1}{N_{c}}\mathcal{C}(\vec{z}) & = & \sum_{\varepsilon_{n}\neq\varepsilon_{v}}\frac{1}{\varepsilon_{n}-\varepsilon_{v}}\phi_{v}^{\dagger}(\vec{z})\phi_{n}(\vec{z})\langle n|\gamma^{0}|v\rangle+\frac{1}{2}\sum_{n,m}\langle n|\gamma^{0}|m\rangle\phi_{m}^{\dagger}(\vec{z})\phi_{n}(\vec{z})\mathcal{R}_{5}(\varepsilon_{n},\varepsilon_{m}),\\
\frac{1}{N_{c}}\mathcal{K}_{1}(\vec{z}) & = & \frac{1}{2}\sum_{\varepsilon_{n}\neq\varepsilon_{v}}\frac{1}{\varepsilon_{n}-\varepsilon_{v}}\langle v|\gamma^{0}\tau^{i}|n\rangle\phi_{n}^{\dagger}(\vec{z})\tau^{i}\phi_{v}(\vec{z})+\frac{1}{4}\sum_{n,m}\langle n|\gamma^{0}\tau^{i}|m\rangle\phi_{m}^{\dagger}(\vec{z})\tau^{i}\phi_{n}(\vec{z})\mathcal{R}_{5}(\varepsilon_{n},\varepsilon_{m}),\\
\frac{1}{N_{c}}\mathcal{K}_{2}(\vec{z}) & = & \frac{1}{4}\sum_{\varepsilon_{n^{0}}}\frac{1}{\varepsilon_{n^{0}}-\varepsilon_{v}}\phi_{v}^{\dagger}(\vec{z})\phi_{n^{0}}(\vec{z})\langle n^{0}|\gamma^{0}|v\rangle+\frac{1}{4}\sum_{n,m}\mathcal{R}_{5}(\varepsilon_{n},\varepsilon_{m^{0}})\phi_{m^{0}}^{\dagger}(\vec{z})\phi_{n}(\vec{z})\langle n|\gamma^{0}|m^{0}\rangle.\end{eqnarray*}
 The vectors $\langle n|$ are eigenstates of the $\chi$QSM Hamiltonian
$h(U)$ which are a linear combination of the eigenstates $\langle n^{0}|$
of the Hamiltonian $H(1)$ \cite{WakamatsuBasis}.

\subsection{$\chi$QSM Magnetic Densities}

The operator for the magnetic form factors in the $\chi$QSM is $O_{1}=\gamma^{0}[\vec{z}\times\vec{\gamma}]_{3}=\gamma^{5}[\vec{z}\times\vec{\sigma}]_{10}$
and the magnetic densities of Eq.(\ref{eq:Mag dens}) are \begin{eqnarray*}
\frac{1}{N_{c}}\mathcal{Q}_{0}(\vec{z}) & = & \langle v||\vec{z}\rangle\{O_{1}\otimes\tau_{1}\}_{0}\langle\vec{z}||v\rangle+\sum_{n}\sqrt{2G_{n}+1}\langle n||\vec{z}\rangle\{O_{1}\otimes\tau_{1}\}_{0}\langle\vec{z}||n\rangle\mathcal{R}_{1}(\varepsilon_{n}),\\
\frac{1}{N_{c}}\mathcal{X}_{1}(\vec{z}) & = & \sum_{\varepsilon_{n}\neq\varepsilon_{v}}\frac{1}{\varepsilon_{n}-\varepsilon_{v}}(-)^{G_{n}}\langle v||\vec{z}\rangle O_{1}\langle\vec{z}||n\rangle\langle n||\tau_{1}||v\rangle\\
 &  & +\frac{1}{2}\sum_{n,m}\mathcal{R}_{5}(\varepsilon_{n},\varepsilon_{m})(-)^{G_{m}-G_{n}}\langle n||\tau_{1}||m\rangle\langle m||\vec{z}\rangle O_{1}\langle\vec{z}||n\rangle,\\
\frac{1}{N_{c}}\mathcal{X}_{2}(\vec{z}) & = & \sum_{\varepsilon_{n^{0}}}\frac{1}{\varepsilon_{n^{0}}-\varepsilon_{v}}\langle n^{0}||\vec{z}\rangle\{O_{1}\otimes\tau_{1}\}_{0}\langle\vec{z}||v\rangle\langle v\mid n^{0}\rangle\\
 &  & +\sum_{n,m^{0}}\mathcal{R}_{5}(\varepsilon_{n},\varepsilon_{m^{0}})\sqrt{2G_{m}+1}\langle m^{0}||\vec{z}\rangle\{O_{1}\otimes\tau_{1}\}_{0}\langle\vec{z}||n\rangle\langle n\mid m^{0}\rangle,\\
\frac{1}{N_{c}}\mathcal{Q}_{1}(\vec{z}) & = & \sum_{\varepsilon_{n}}\frac{\textrm{sign}(\varepsilon_{n})}{\varepsilon_{n}-\varepsilon_{v}}(-)^{G_{n}}\langle n||\vec{z}\rangle\{O_{1}\otimes\tau_{1}\}_{1}\langle\vec{z}||v\rangle\langle v||\tau_{1}||n\rangle\\
 &  & +\frac{1}{2}\sum_{n,m}\mathcal{R}_{4}(\varepsilon_{n},\varepsilon_{m})(-)^{G_{m}-G_{n}}\langle n||\vec{z}\rangle\{O_{1}\otimes\tau_{1}\}_{1}\langle\vec{z}||m\rangle\langle m||\tau_{1}||n\rangle,\\
\frac{1}{N_{c}}\mathcal{M}_{0}(\vec{z}) & = & \sum_{\varepsilon_{n}\neq\varepsilon_{v}}\frac{1}{\varepsilon_{n}-\varepsilon_{v}}\langle v||\vec{z}\rangle\{O_{1}\otimes\tau_{1}\}_{0}\langle\vec{z}||n\rangle\langle n|\gamma^{0}|v\rangle\\
 &  & -\frac{1}{2}\sum_{n,m}\mathcal{R}_{2}(\varepsilon_{n},\varepsilon_{m})\sqrt{2G_{m}+1}\langle n|\gamma^{0}|m\rangle\langle m||\vec{z}\rangle\{O_{1}\otimes\tau_{1}\}_{0}\langle\vec{z}||n\rangle,\\
\frac{1}{N_{c}}\mathcal{M}_{1}(\vec{z}) & = & \sum_{\varepsilon_{n}\neq\varepsilon_{v}}\frac{1}{\varepsilon_{n}-\varepsilon_{v}}(-)^{G_{n}}\langle n||\gamma^{0}\tau_{1}||v\rangle\langle v||\vec{z}\rangle O_{1}\langle\vec{z}||n\rangle\\
 &  & -\frac{1}{2}\sum_{n,m}\mathcal{R}_{2}(\varepsilon_{n},\varepsilon_{m})(-)^{G_{m}-G_{n}}\langle n||\gamma^{0}\tau_{1}||m\rangle\langle m||\vec{z}\rangle O_{1}\langle\vec{z}||n\rangle,\\
\frac{1}{N_{c}}\mathcal{M}_{2}(\vec{z}) & = & \sum_{\varepsilon_{n^{0}}}\frac{1}{\varepsilon_{n^{0}}-\varepsilon_{v}}\langle v||\vec{z}\rangle\{O_{1}\otimes\tau_{1}\}_{0}\langle\vec{z}||n^{0}\rangle\langle n^{0}|\gamma^{0}|v\rangle\\
 &  & -\sum_{n,m^{0}}\mathcal{R}_{2}(\varepsilon_{n},\varepsilon_{m^{0}})\sqrt{2G_{m}+1}\langle m^{0}||\vec{z}\rangle\{O_{1}\otimes\tau_{1}\}_{0}\langle\vec{z}||n\rangle\langle n|\gamma^{0}|m^{0}\rangle.\end{eqnarray*}

\subsection{Regularization Functions}

The regularization functions are defined as: \begin{eqnarray}
\mathcal{R}_{1}(\varepsilon_{n}) & = & -\frac{1}{2\sqrt{\pi}}\varepsilon_{n}\int_{1/\Lambda^{2}}^{\infty}\frac{du}{\sqrt{u}}e^{-u\varepsilon_{n}^{2}},\\
\mathcal{R}_{2}(\varepsilon_{n},\varepsilon_{m}) & = & \int_{1/\Lambda^{2}}^{\infty}du\frac{1}{2\sqrt{\pi u}}\frac{\varepsilon_{m}e^{-u\varepsilon_{m}^{2}}-\varepsilon_{n}e^{-u\varepsilon_{n}^{2}}}{\varepsilon_{n}-\varepsilon_{m}},\\
\mathcal{R}_{3}(\varepsilon_{n},\varepsilon_{m}) & = & \frac{1}{2\sqrt{\pi}}\int_{1/\Lambda^{2}}^{\infty}\frac{du}{\sqrt{u}}\Big[\frac{1}{u}\frac{e^{-\varepsilon_{n}^{2}u}-e^{-\varepsilon_{m}^{2}u}}{\varepsilon_{m}^{2}-\varepsilon_{n}^{2}}-\frac{\varepsilon_{n}e^{-u\varepsilon_{n}^{2}}+\varepsilon_{m}e^{-u\varepsilon_{m}^{2}}}{\varepsilon_{m}+\varepsilon_{n}}\Big],\\
\mathcal{R}_{4}(\varepsilon_{n},\varepsilon_{m}) & = & \frac{1}{2\pi}\int_{1/\Lambda^{2}}^{\infty}du\int_{0}^{1}d\alpha e^{-\varepsilon_{n}^{2}u(1-\alpha)-\alpha\varepsilon_{m}^{2}u}\frac{\varepsilon_{n}(1-\alpha)-\alpha\varepsilon_{m}}{\sqrt{\alpha(1-\alpha)}},\\
\mathcal{R}_{5}(\varepsilon_{n},\varepsilon_{m}) & = & \frac{1}{2}\frac{\textrm{sign}\varepsilon_{n}-\textrm{sign}\varepsilon_{m}}{\varepsilon_{n}-\varepsilon_{m}},\\
\mathcal{R}_{6}(\varepsilon_{n},\varepsilon_{m}) & = & \frac{1-\textrm{sign}(\varepsilon_{n})\textrm{sign}(\varepsilon_{m})}{\varepsilon_{n}-\varepsilon_{m}}.\end{eqnarray}

\subsection{\label{sub:APP redmatelem}Reduced Matrix Elements for $\{\sqrt{4\pi}Y_{2}\otimes\tau_{1}\}_{1}$}

We use the basis of \cite{WakamatsuBasis} where the iso-spin $\tau$
and total angular momentum $j$ is coupled to the grand-spin $G=\tau+j$
($j=l+s$)\begin{eqnarray}
|0\rangle & = & |l=G\,\,;\,\, j=G+\frac{1}{2}\,\,;\,\, GG_{3}\rangle\,\,\,,\\
|1\rangle & = & |l=G\,\,;\,\, j=G-\frac{1}{2}\,\,;\,\, GG_{3}\rangle\,\,\,,\\
|2\rangle & = & |l=G+1\,\,;\,\, j=G+\frac{1}{2}\,\,;\,\, GG_{3}\rangle\,\,\,,\\
|3\rangle & = & |l=G-1\,\,;\,\, j=G-\frac{1}{2}\,\,;\,\, GG_{3}\rangle\,\,\,.\end{eqnarray}
 The reduced matrix elements for the operator $\{\sqrt{4\pi}Y_{2}\otimes\tau_{1}\}_{1}$
in the density $\mathcal{I}_{1E2}(r)$ Eq.(\ref{eq:GE2 CQSM Final})
are with the notation $\langle n||\{\sqrt{4\pi}Y_{2}\otimes\tau_{1}\}_{1}||m\rangle$:

\begin{minipage}[t][1\totalheight]{0.49\columnwidth}%
\begin{eqnarray*}
A^{0}(G) & = & (-)(G+2)\sqrt{\frac{2G}{(2G+1)(G+1)}}\\
B^{0}(G) & = & (-)3\sqrt{\frac{1}{2(2G+1)}}\\
C^{0}(G) & = & (G-1)\sqrt{\frac{(2G+2)}{(G)(2G+1)}}\\
\\\end{eqnarray*}
\end{minipage}%
\begin{minipage}[t][1\totalheight]{0.49\columnwidth}%
\begin{eqnarray*}
A^{1}(G) & = & (-)G(2G+4)\sqrt{\frac{1}{(2G+1)(2G+2)(2G+3)}}\\
B^{1}(G) & = & 3\sqrt{\frac{(G+2)}{2(2G+1)(2G+3)}}\\
C^{1}(G) & = & (-)3\sqrt{\frac{G(G+1)(2G+4)}{(2G+1)(2G+3)}}\\
D^{1}(G) & = & (-)3\sqrt{\frac{G}{2(2G+1)(2G+3)}}\end{eqnarray*}
\end{minipage}\\
 \\
 \\
 \begin{tabular}{|c|cccc|}
\hline 
$G^{m}=G^{n}$  & $|0(G)\rangle$  & $|1(G)\rangle$  & $|2(G)\rangle$  & $|3(G)\rangle$\tabularnewline
\hline
\hline 
$\langle0(G)|$  & $A^{0}(G)$  & $B^{0}(G)$  & $0$  & $0$\tabularnewline
$\langle1(G)|$  & $B^{0}(G)$  & $C^{0}(G)$  & $0$  & $0$\tabularnewline
$\langle2(G)|$  & $0$  & $0$  & $A^{0}(G)$  & $B^{0}(G)$\tabularnewline
$\langle3(G)|$  & $0$  & $0$  & $B^{0}(G)$  & $C^{0}(G)$\tabularnewline
\hline
\end{tabular}~~~\begin{tabular}{|c|cccc|}
\hline 
$G^{m}=G^{n}+1$  & $|0(G+1)\rangle$  & $|1(G+1)\rangle$  & $|2(G+1)\rangle$  & $|3(G+1)\rangle$\tabularnewline
\hline
\hline 
$\langle0(G)|$  & $0$  & $0$  & $B^{1}(G)$  & $A^{1}(G)$\tabularnewline
$\langle1(G)|$  & $0$  & $0$  & $C^{1}(G)$  & $D^{1}(G)$\tabularnewline
$\langle2(G)|$  & $B^{1}(G)$  & $A^{1}(G)$  & $0$  & $0$\tabularnewline
$\langle3(G)|$  & $C^{1}(G)$  & $D^{1}(G)$  & $0$  & $0$\tabularnewline
\hline
\end{tabular}

\subsection{Matrix-Elements\label{sub:Matrix-Elements}}

The baryon matrix-elements, such as $\langle B^{\prime}|D_{\chi3}^{(8)}|B\rangle$,
are evaluated by using the $SU(3)$ group algebra \cite{CHP,deSwart}
\begin{eqnarray}
\langle B_{\mathcal{R}^{\prime}}^{\prime}|D_{\chi m}^{n}(A)|B_{\mathcal{R}}\rangle & = & \sqrt{\frac{\textrm{dim}\mathcal{R}^{\prime}}{\textrm{dim}\mathcal{R}}}(-1)^{\frac{1}{2}Y_{s}^{\prime}+S_{3}^{\prime}}(-1)^{\frac{1}{2}Y_{s}+S_{3}}\nonumber \\
 &  & \times\sum_{\gamma}\left(\begin{array}{ccc}
\mathcal{R}^{\prime} & n & \mathcal{R}_{\gamma}\\
Q^{\prime} & \chi & Q\end{array}\right)\left(\begin{array}{ccc}
\mathcal{R}^{\prime} & n & \mathcal{R}_{\gamma}\\
-Y_{s}^{\prime}S^{\prime}-S_{3}^{\prime} & m & -Y_{S}S-S_{3}\end{array}\right)\,\,\,,\end{eqnarray}
 with $Q=YII_{3}$. $\left(\cdots\right)$ denote the $SU(3)$ Clebsch-Gordan
coefficients.\\
 The wave function corrections Eq.(\ref{wfc}) for the other decuplet
baryons are

\begin{eqnarray}
|B_{10}\rangle & = & |10_{3/2},B\rangle+a_{27}^{B}|27_{3/2},B\rangle+a_{35}^{B}|35_{3/2},B\rangle\,\,\,,\end{eqnarray}
 with the mixing coefficients \begin{eqnarray}
a_{27}^{B} & = & a_{27}\left(\begin{array}{c}
\sqrt{15/2}\\
2\\
\sqrt{3/2}\\
0\end{array}\right),\;\; a_{35}^{B}=a_{35}\left(\begin{array}{c}
5/\sqrt{14}\\
2\,\sqrt{5/7}\\
3\,\sqrt{5/14}\\
2\,\sqrt{5/7}\end{array}\right)\,\,\,,\end{eqnarray}
 in the bases $[\Delta,\Sigma_{10}^{*},\Xi_{10}^{*},\Omega]$.

\subsubsection{Magnetic Part}

We take the abbreviation $d_{ab3}D_{\chi b}^{(8)}J_{a}=dD_{\chi}J$
and the matrix-element for the magnetic form factors of the decuplet
baryons read as follows with $|B_{10}\rangle=|B_{10}(Y,I_{3},S_{3})\rangle$:\\
 \\
 \textbf{Leading order}:\\
 \\
 \begin{tabular}{cc}
\hline 
{\tiny $\langle\Delta|D_{33}^{(8)}|\Delta\rangle=\langle\Sigma_{10}|D_{33}^{(8)}|\Sigma_{10}\rangle=\langle\Xi_{10}|D_{33}^{(8)}|\Xi_{10}\rangle=\langle\Omega|D_{33}^{(8)}|\Omega\rangle=$}  & {\tiny $-I_{3}\frac{1}{6}S_{3}$}\tabularnewline
{\tiny $\langle\Delta|D_{38}^{(8)}S_{3}|\Delta\rangle=\langle\Sigma_{10}|D_{38}^{(8)}S_{3}|\Sigma_{10}\rangle=\langle\Xi_{10}|D_{38}^{(8)}S_{3}|\Xi_{10}\rangle=\langle\Omega|D_{38}^{(8)}S_{3}|\Omega\rangle=$}  & {\tiny $I_{3}\frac{1}{4}\sqrt{\frac{1}{3}}S_{3}$}\tabularnewline
{\tiny $\langle\Delta|dD_{3}J|\Delta\rangle=\langle\Sigma_{10}|dD_{3}J|\Sigma_{10}\rangle=\langle\Xi_{10}|dD_{3}J|\Xi_{10}\rangle=\langle\Omega|dD_{3}J|\Omega\rangle=$}  & {\tiny $I_{3}\frac{1}{12}S_{3}$}\tabularnewline
\hline 
{\tiny $\langle\Delta|D_{83}^{(8)}|\Delta\rangle=\langle\Sigma_{10}|D_{83}^{(8)}|\Sigma_{10}\rangle=\langle\Xi_{10}|D_{83}^{(8)}|\Xi_{10}\rangle=\langle\Omega|D_{83}^{(8)}|\Omega\rangle=$}  & {\tiny $-Y\frac{1}{4}\sqrt{\frac{1}{3}}S_{3}$}\tabularnewline
{\tiny $\langle\Delta|D_{88}^{(8)}S_{3}|\Delta\rangle=\langle\Sigma_{10}|D_{88}^{(8)}S_{3}|\Sigma_{10}\rangle=\langle\Xi_{10}|D_{88}^{(8)}S_{3}|\Xi_{10}\rangle=\langle\Omega|D_{88}^{(8)}S_{3}|\Omega\rangle=$}  & {\tiny $Y\frac{1}{8}S_{3}$}\tabularnewline
{\tiny $\langle\Delta|dD_{8}J|\Delta\rangle=\langle\Sigma_{10}^{*}|dD_{8}J|\Sigma_{10}^{*}\rangle=\langle\Xi_{10}^{*}|dD_{8}J|\Xi_{10}^{*}\rangle=\langle\Omega|dD_{8}J|\Omega\rangle=$}  & {\tiny $Y\frac{1}{8}\frac{1}{\sqrt{3}}S_{3}$}\tabularnewline
\hline
\end{tabular}\\
 \\
 \textbf{Wavefunction corrections:} \\
 \\
 \begin{tabular}{c}
{\tiny $\langle\Omega|D_{33}^{(8)}|\Omega\rangle=\langle\Omega|D_{38}^{(8)}J_{3}|\Omega\rangle=\langle\Omega|dD_{3}J|\Omega\rangle=0$}\tabularnewline
{\tiny $\langle\Omega|D_{83}^{(8)}|\Omega\rangle=-a_{35}^{B}S_{3}\sqrt{\frac{1}{105}}\,\,\,\,\,\langle\Omega|D_{88}^{(8)}S_{3}|\Omega\rangle=S_{3}a_{35}^{B}\frac{5}{2}\sqrt{\frac{1}{35}}\,\,\,\,\,\langle\Omega|dD_{8}J|\Omega\rangle=-\frac{5}{2}S_{3}\sqrt{\frac{1}{105}}a_{35}^{B}$}\tabularnewline
\end{tabular}\\
 \\
 \\
 \begin{tabular}{c|c|c|c}
\hline 
{\tiny $ $}  & {\tiny $\Delta$}  & {\tiny $ $}  & {\tiny $\Delta$}\tabularnewline
\hline 
{\tiny $D_{33}^{(8)}$}  & {\tiny $I_{3}S_{3}\Big[-a_{27}^{B}\frac{5}{9}\sqrt{\frac{1}{30}}-a_{35}^{B}\frac{1}{15}\sqrt{\frac{1}{14}}\Big]$}  & {\tiny $D_{83}^{(8)}$}  & {\tiny $S_{3}\Big[a_{27}^{B}\frac{5}{6}\sqrt{\frac{1}{10}}-a_{35}^{B}\frac{1}{2}\sqrt{\frac{1}{42}}\Big]$}\tabularnewline
{\tiny $D_{38}^{(8)}J_{3}$}  & {\tiny $I_{3}S_{3}\Big[-a_{27}^{B}\frac{5}{6}\sqrt{\frac{1}{10}}+a_{35}^{B}\frac{1}{2}\sqrt{\frac{1}{42}}\Big]$}  & {\tiny $D_{88}^{(8)}J_{3}$}  & {\tiny $S_{3}\Big[a_{27}^{B}\frac{15}{4}\sqrt{\frac{1}{30}}+a_{35}^{B}\frac{5}{4}\sqrt{\frac{1}{14}}\Big]$}\tabularnewline
{\tiny $dD_{3}J$}  & {\tiny $I_{3}S_{3}\Big[-\frac{5}{18}\sqrt{\frac{1}{30}}a_{27}^{B}-\frac{1}{6}a_{35}^{B}\sqrt{\frac{1}{14}}\Big]$}  & {\tiny $dD_{8}J$}  & {\tiny $S_{3}\Big[\frac{5}{12}\sqrt{\frac{1}{10}}a_{27}^{B}-\frac{5}{4}\sqrt{\frac{1}{42}}a_{35}^{B}\Big]$}\tabularnewline
\hline 
 & {\tiny $\Sigma_{10}^{*}$}  &  & {\tiny $\Sigma_{10}^{*}$}\tabularnewline
\hline 
{\tiny $D_{33}^{(8)}$}  & {\tiny $I_{3}S_{3}\Big[-a_{27}^{B}\frac{1}{6}-a_{35}^{B}\frac{1}{6}\sqrt{\frac{1}{35}}\Big]$}  & {\tiny $D_{83}^{(8)}$}  & {\tiny $S_{3}\Big[a_{27}^{B}\frac{1}{3}\sqrt{\frac{1}{3}}-a_{35}^{B}\sqrt{\frac{1}{105}}\Big]$}\tabularnewline
{\tiny $D_{38}^{(8)}S_{3}$}  & {\tiny $I_{3}S_{3}\Big[-a_{27}^{B}\frac{3}{2}\sqrt{\frac{1}{3}}+a_{35}^{B}\frac{5}{2}\sqrt{\frac{1}{105}}\Big]$}  & {\tiny $D_{88}^{(8)}S_{3}$}  & {\tiny $S_{3}\Big[a_{27}^{B}\frac{1}{2}+a_{35}^{B}\frac{5}{2}\sqrt{\frac{1}{35}}\Big]$}\tabularnewline
{\tiny $dD_{3}J$}  & {\tiny $I_{3}S_{3}\Big[-\frac{1}{12}a_{27}^{B}-\frac{5}{12}\sqrt{\frac{1}{35}}a_{35}^{B}\Big]$}  & {\tiny $dD_{8}J$}  & {\tiny $S_{3}\Big[\frac{1}{6}a_{27}^{B}\sqrt{\frac{1}{3}}-\frac{5}{2}\sqrt{\frac{1}{105}}a_{35}^{B}\Big]$}\tabularnewline
\hline 
 & {\tiny $\Xi_{10}^{*}$}  &  & {\tiny $\Xi_{10}^{*}$}\tabularnewline
\hline 
{\tiny $D_{33}^{(8)}$}  & {\tiny $I_{3}S_{3}\Big[-a_{27}^{B}\frac{7}{9}\sqrt{\frac{1}{6}}-a_{35}^{B}\frac{1}{3}\sqrt{\frac{1}{70}}\Big]$}  & {\tiny $D_{83}^{(8)}$}  & {\tiny $S_{3}\Big[a_{27}^{B}\frac{1}{6}\sqrt{\frac{1}{2}}-a_{35}^{B}\frac{3}{2}\sqrt{\frac{1}{210}}\Big]$}\tabularnewline
{\tiny $D_{38}^{(8)}S_{3}$}  & {\tiny $I_{3}S_{3}\Big[-a_{27}^{B}\frac{7}{6}\sqrt{\frac{1}{2}}+a_{35}^{B}\frac{5}{2}\sqrt{\frac{1}{210}}\Big]$}  & {\tiny $D_{88}^{(8)}S_{3}$}  & {\tiny $S_{3}\Big[a_{27}^{B}\frac{3}{4}\sqrt{\frac{1}{6}}+a_{35}^{B}\frac{15}{4}\sqrt{\frac{1}{70}}\Big]$}\tabularnewline
{\tiny $dD_{3}J$}  & {\tiny $I_{3}S_{3}\Big[-\frac{7}{18}a_{27}^{B}\sqrt{\frac{1}{6}}-\frac{5}{6}\sqrt{\frac{1}{70}}a_{35}^{B}\Big]$}  & {\tiny $dD_{8}J$}  & {\tiny $S_{3}\Big[\frac{1}{12}\sqrt{\frac{1}{2}}a_{27}^{B}-\frac{15}{4}\sqrt{\frac{1}{210}}a_{35}^{B}\Big]$}\tabularnewline
\hline
\end{tabular}\\
 \\
 \\
 \textbf{Operator corrections}~$D_{88}^{(8)}D_{83}^{(8)}=D_{83}^{(8)}D_{88}^{(8)}$\\
 \\
 \begin{tabular}{c|cccc}
\hline 
 & {\tiny $\Delta$}  & {\tiny $\Sigma_{10}^{*}$}  & {\tiny $\Xi_{10}^{*}$}  & {\tiny $\Omega$}\tabularnewline
\hline 
{\tiny $D_{88}^{(8)}D_{33}^{(8)}$}  & {\tiny $-S_{3}I_{3}\frac{5}{126}$}  & {\tiny $-S_{3}I_{3}\frac{1}{42}$}  & {\tiny $-S_{3}I_{3}\frac{1}{126}$}  & {\tiny $0$}\tabularnewline
{\tiny $D_{83}^{(8)}D_{38}^{(8)}$}  & {\tiny $-S_{3}I_{3}\frac{5}{126}$}  & {\tiny $-I_{3}\frac{1}{42}$}  & {\tiny $-S_{3}I_{3}\frac{1}{126}$}  & {\tiny $0$}\tabularnewline
{\tiny $D_{8a}^{(8)}D_{3b}^{(8)}d_{ab3}$}  & {\tiny $-S_{3}I_{3}\frac{11}{126}\sqrt{\frac{1}{3}}$}  & {\tiny $-S_{3}I_{3}\frac{5}{42}\sqrt{\frac{1}{3}}$}  & {\tiny $-S_{3}I_{3}\frac{19}{126}\sqrt{\frac{1}{3}}$}  & {\tiny $0$}\tabularnewline
{\tiny $D_{83}^{(8)}D_{88}^{(8)}$}  & {\tiny $S_{3}\frac{1}{28}\sqrt{\frac{1}{3}}$}  & {\tiny $S_{3}\frac{1}{42}\sqrt{\frac{1}{3}}$}  & {\tiny $-S_{3}\frac{1}{28}\sqrt{\frac{1}{3}}$}  & {\tiny $-S_{3}\frac{1}{7}\sqrt{\frac{1}{3}}$}\tabularnewline
{\tiny $D_{8a}^{(8)}D_{8b}^{(8)}d_{ab3}$}  & {\tiny $S_{3}\frac{5}{84}$}  & {\tiny $-S_{3}\frac{1}{63}$}  & {\tiny $-S_{3}\frac{5}{84}$}  & {\tiny $-S_{3}\frac{1}{14}$}\tabularnewline
\hline
\end{tabular}

\subsubsection{Electric Part}

The $\Delta$ state $|\Delta\rangle$ is explicitly $|\Delta\rangle=|\Delta(I_{3},S_{3})\rangle$
and the matrix elements for the electric form factor read:\\
\begin{minipage}[t][1\totalheight]{0.49\columnwidth}%
\begin{eqnarray*}
\langle\Delta|D_{38}^{8}|\Delta\rangle & = & I_{3}\Big[\frac{1}{4}\sqrt{\frac{1}{3}}-a_{27}^{\Delta}\frac{5}{6}\sqrt{\frac{1}{10}}+a_{35}^{\Delta}\frac{1}{2}\sqrt{\frac{1}{42}}\Big]\\
\langle\Delta|D_{3i}^{(8)}J_{i}|\Delta\rangle & = & I_{3}\Big[-\frac{5}{8}-a_{27}^{\Delta}\frac{25}{12}\sqrt{\frac{1}{30}}-a_{35}^{\Delta}\frac{1}{4}\sqrt{\frac{1}{14}}\Big]\\
\langle\Delta|D_{3a}^{(8)}J_{a}|\Delta\rangle & = & I_{3}\Big[-\frac{1}{4}+a_{27}^{\Delta}\frac{5}{6}\,\sqrt{\frac{1}{30}}+a_{35}^{\Delta}\frac{1}{2}\sqrt{\frac{1}{14}}\Big]\\
\langle\Delta|D_{88}^{(8)}|\Delta\rangle & = & \frac{1}{8}+a_{27}^{\Delta}\frac{15}{4}\sqrt{\frac{1}{30}}+a_{35}^{\Delta}\frac{5}{4}\sqrt{\frac{1}{14}}\\
\langle\Delta|D_{8i}^{(8)}J_{i}|\Delta\rangle & = & -\frac{15}{16}\sqrt{\frac{1}{3}}+a_{27}^{\Delta}\frac{25}{8}\sqrt{\frac{1}{10}}-a_{35}^{\Delta}\frac{15}{8}\sqrt{\frac{1}{42}}\\
\langle\Delta|D_{8a}^{(8)}J_{a}|\Delta\rangle & = & -\frac{3}{8}\frac{1}{\sqrt{3}}-a_{27}^{\Delta}\frac{5}{4}\sqrt{\frac{1}{10}}+a_{35}^{\Delta}\frac{15}{4}\sqrt{\frac{1}{42}}\end{eqnarray*}
\end{minipage}%
\begin{minipage}[t][1\totalheight]{0.49\columnwidth}%
\begin{eqnarray*}
\langle\Delta|D_{8i}^{(8)}D_{3i}^{(8)}|\Delta\rangle & = & I_{3}\frac{13}{84}\sqrt{\frac{1}{3}}\\
\langle\Delta|D_{8a}^{(8)}D_{3a}^{(8)}|\Delta\rangle & = & -I_{3}\frac{5}{42}\sqrt{\frac{1}{3}}\\
\langle\Delta|D_{88}^{(8)}D_{38}^{(8)}|\Delta\rangle & = & -I_{3}\frac{1}{28}\sqrt{\frac{1}{3}}\\
\langle\Delta|D_{8i}^{(8)}D_{8i}^{(8)}|\Delta\rangle & = & \frac{17}{56}\\
\langle\Delta|D_{8a}^{(8)}D_{8a}^{(8)}|\Delta\rangle & = & \frac{15}{28}\\
\langle\Delta|D_{88}^{(8)}D_{88}^{(8)}|\Delta\rangle & = & \frac{9}{56}\end{eqnarray*}
\end{minipage}

\end{document}